\documentclass[lettersize,journal]{IEEEtran}
\IEEEoverridecommandlockouts

\usepackage{color}

\definecolor{AlirezaPurple}{RGB}{150, 0, 250}

\usepackage{cite}
\usepackage{amsmath,amssymb,amsfonts}
\usepackage{algorithm}
\usepackage{algorithmic}
\usepackage{acronym}
\usepackage{tikz}
\usepackage{graphicx}
\usepackage{textcomp}
\usepackage{xcolor}
\usepackage{pgfplots}
\pgfplotsset{compat=1.18}
\def\BibTeX{{\rm B\kern-.05em{\sc i\kern-.025em b}\kern-.08em
    T\kern-.1667em\lower.7ex\hbox{E}\kern-.125emX}}
\usepackage{titlesec}
\usepackage{geometry}
\acrodef{ad}[AD]{autonomous drive}
\acrodef{alb}[ALB]{absolute lower bound}
\acrodef{adas}[ADAS]{advanced driver assistance system}
\acrodef{aoa}[AOA]{angles-of-arrival}
\acrodef{aod}[AOD]{angles-of-departure}
\acrodef{bs}[BS]{base station}
\acrodef{brs}[BRS]{beam response similarity}
\acrodef{cdf}[CDF]{cumulative distribution function}
\acrodef{crb}[CRB]{Cram\'er–Rao bound}
\acrodef{dbscan}[DBSCAN]{density-based spatial clustering of applications with noise}
\acrodef{gnss}[GNSS]{global navigation satellite system}
\acrodef{gps}[GPS]{global positioning system}
\acrodef{idft}[IDFT]{inverse discrete Fourier transform}
\acrodef{imu}[IMU]{inertial measurement unit}
\acrodef{isac}[ISAC]{integrated sensing and communication}
\acrodef{ip}[IP]{incidence point}
\acrodef{las}[L\&S]{localization and sensing}
\acrodef{los}[LOS]{line-of-sight}
\acrodef{mae}[MAE]{mean absolute value}
\acrodef{mc}[MC]{mutual coupling}
\acrodef{mcrb}[MCRB]{misspecified Cram\'er–Rao bound}
\acrodef{mcm}[MCM]{mutual coupling matrix}
\acrodef{map}[MAP]{maximum a posteriori}
\acrodef{mle}[MLE]{maximum likelihood estimator}
\acrodef{mpc}[MPC]{multipath component}
\acrodef{nlos}[NLOS]{non-line-of-sight}
\acrodef{nc}[NC]{non-ideal codebook}
\acrodef{ci}[CI]{combined impacts}
\acrodef{omp}[OMP]{orthogonal matching pursuit}
\acrodef{ofdm}[OFDM]{orthogonal frequency division multiplexing}
\acrodef{prs}[PRS]{positioning reference signal}
\acrodef{pdf}[PDF]{probability density function}
\acrodef{pss}[PSS]{primary synchronization signal}
\acrodef{rf}[RF]{radio frequency}
\acrodef{ris}[RIS]{reconfigurable intelligent surface}
\acrodef{rev}[REV]{rotating element electric field vector}
\acrodef{rss}[RSS]{received signal strength}
\acrodef{rcs}[RCS]{radar cross section}
\acrodef{rtk}[RTK]{real-time kinematic}
\acrodef{rtt}[RTT]{round-trip-time}
\acrodef{slam}[SLAM]{simultaneous localization and mapping}
\acrodef{sp}[SP]{scattering point}
\acrodef{ssb}[SSB]{synchronization signal/physical broadcast channel block}
\acrodef{siso}[SISO]{single-input-single-output}
\acrodef{snr}[SNR]{signal-to-noise ratio}
\acrodef{tdoa}[TDOA]{time-difference-of-arrival}
\acrodef{toa}[TOA]{time-of-arrival}
\acrodef{ue}[UE]{user equipment}
\acrodef{ura}[URA]{uniform rectangular array}
\acrodef{va}[VA]{virtual anchor}
\acrodef{em}[EM]{electromagnetic}

\long\def\comment#1{}

\newfont{\bbb}{msbm10 scaled 700}

\newfont{\bb}{msbm10 scaled 1100}

\newcommand{\av}{{\bf a}}
\newcommand{\bv}{{\bf b}}

\newcommand{\dv}{{\bf d}}

\newcommand{\gv}{{\bf g}}

\newcommand{\ov}{{\bf o}}
\newcommand{\pv}{{\bf p}}

\newcommand{\sv}{{\bf s}}

\newcommand{\wv}{{\bf w}}

\newcommand{\yv}{{\bf y}}

\newcommand{\Am}{{\bf A}}
\newcommand{\Bm}{{\bf B}}

\newcommand{\Qm}{{\bf Q}}
\newcommand{\Rm}{{\bf R}}

\newcommand{\Wm}{{\bf W}}

\newcommand{\Xm}{{\bf X}}
\newcommand{\Ym}{{\bf Y}}

\newcommand{\etav}{\hbox{\boldmath$\eta$}}

\newcommand{\epsilonv}{\hbox{\boldmath$\epsilon$}}

\newcommand{\muv}{\hbox{\boldmath$\mu$}}

\newcommand{\thetav}{\hbox{$\boldsymbol\theta$}}

\newcommand{\xiv}{\hbox{\boldmath$\xi$}}

\newcommand{\varphiv}{\hbox{\boldmath$\varphi$}}

\newcommand{\Gammam}{\hbox{\boldmath$\Gamma$}}

\newcommand{\Phim}{\hbox{\boldmath$\Phi$}}

\newcommand{\trace}{{\hbox{tr}}}

\renewcommand{\arg}{{\hbox{arg}}}

\newcommand{\herm}{{\sf H}}

\begin{document}

\bstctlcite{IEEEexample:BSTcontrol}

\title{On-Site Beam Calibration for RIS-Aided Positioning Systems}

\author{
Mengting~Li,~\IEEEmembership{Member,~IEEE},
Hui~Chen,~\IEEEmembership{Member,~IEEE}, 
Sigurd S. Petersen,
Alireza~Pourafzal,~\IEEEmembership{Member,~IEEE}, 
Huiping~Huang,~\IEEEmembership{Member,~IEEE}, 
Ming~Shen,~\IEEEmembership{Senior Member,~IEEE},
Mikko~Valkama,~\IEEEmembership{Fellow,~IEEE},
Henk~Wymeersch,~\IEEEmembership{Fellow,~IEEE}
\thanks{M.~Li, H.~Chen, A.~Pourafzal, and H.~Wymeersch are with the Department of Electrical Engineering, Chalmers University of Technology, 412 58 Gothenburg, Sweden (Email: {limeng@chalmers.se, hui.chen, alireza.pourafzal, henkw}@chalmers.se). }
\thanks{H.~Huang is with the Department of Engineering Science, Macau University of Science and Technology, 999078 Taipa, Macau SAR, China (Email: hphuang@must.edu.mo).}
\thanks{M.~Li, S. S. Petersen and M. Shen are with the Aalborg University, 9220 Aalborg, Denmark. (Email: {mengli; mish}@es.aau.dk, sp19@student.aau.dk)}
\thanks{M.~Valkama is with the Department of Electrical Engineering, Tampere University, 33100 Tampere, Finland. (Email: mikko.valkama@tuni.fi).}

\thanks{This work was supported, in part, by the research grant (VIL59841) from VILLUM FONDEN, the Swedish Research Council (VR grant 2023-03821), the SNS JU project 6G-DISAC under the EU's Horizon Europe Research and Innovation Program under Grant Agreement No. 101139130, and the European Union's Horizon Europe research and innovation programme under the Marie Sk\l{}odowska-Curie grant agreement No. 101207620 (MIDAS-6G) and No. 101201808 (SMALL-6G).}
}
\maketitle
\begin{abstract}
High precision positioning is a key enabler for next-generation communication applications such as smart transportation and augmented reality. Reconfigurable intelligent surface (RIS) technology can enhance positioning by providing additional angular information and improving coverage under obstructed propagation conditions. However, true RIS beams can differ significantly from the simplified or ideal beam response models commonly used in RIS-aided positioning, leading to beam model mismatch and an elevated positioning error floor. This paper proposes an on-site RIS beam calibration framework that reduces this error floor by estimating a realistic 3D RIS beam response model from on-site measurements. The proposed calibration algorithm first extracts the RIS-reflected channel response from signals received by a calibration agent sampling the angular range of interest, using delay-domain sparse recovery, and then estimates the beam model parameters with a gradient-based estimator. To validate the proposed framework, 3D beam patterns under 66 phase modulations were measured and incorporated into simulations. With an angular sampling step of $1^\circ$, the calibrated model achieves an average beam response similarity of 88.5\% with respect to the ground truth, compared with 43.7\% for the ideal model. The probability that the absolute lower bound of the positioning error is below $0.5~\mathrm{m}$ increases from 0.52 without calibration to 0.74 after calibration, showing that on-site RIS beam calibration effectively reduces the positioning error floor caused by true beam model mismatch.
\end{abstract}


\section{Introduction}\label{sec:intro}
Positioning in fifth- and sixth-generation (5G/6G) radio systems has attracted significant attention from both academia and industry, as the use of wide bandwidth and large-scale antenna arrays unlocks the potential for high precision positioning especially in environments such as indoor settings and urban canyons where global navigation satellite systems (GNSS) face limitations~\cite{gao2023dl}. High precision positioning enables future radio systems to offer location-aware, context-aware, and environment-aware services, in addition to delivering high data rates and broad connectivity. \Ac{ris} has been widely adopted in positioning systems due to its cost-effectiveness and ease of deployment. With its ability to manipulate \ac{em} waves, an \ac{ris} serving as a passive anchor, can enhance both the reliability~\cite{liu2021reconfigurable} and accuracy~\cite{chen2024multi} of positioning systems across a variety of scenarios.

Many existing studies on RIS-aided positioning typically use an ideal array model to compute the \ac{ris} beam response, assuming isotropic element patterns, ideal phase tuning while neglecting hardware impairments and array imperfections~\cite{chen2024multi,keykhosravi2022ris}. However, the true \ac{ris} beam response will be altered from the ideal model mainly due to non-ideal phase tuning networks, mutual coupling, and environmental effects. Non-ideal phase tuning includes effects such as changes in amplitude of reflection coefficients and discrepancies between actual and intended phase responses. Mutual coupling, commonly observed in antenna arrays and \ac{ris}, results in the interdependence of element responses influenced by their surrounding \ac{em} environment~\cite{yuan2023effects}. Ultimately, these mutual coupling effects will also cause deviations in phase tuning and variations in element patterns (i.e., the reflection coefficients across different angles of incidence for each RIS element)~\cite{chen2018review}. Besides, environmental conditions such as humidity, temperature, as well as fabrication errors can alter or cause drift in the \ac{em} characteristics of the RIS elements. Neglecting these factors in the positioning algorithm can lead to a mismatch between the assumed and actual \ac{ris} beam responses, thereby elevating the positioning error floor~\cite{li2025ris}. Therefore, calibration is required to reduce model mismatch and ensure high precision positioning. 

Calibration of antenna arrays has been widely explored, whereas fewer studies have focused on RIS calibration, especially for supporting positioning and sensing. Calibration in communication systems primarily focuses on compensating excitation (for arrays) or reflection coefficient (for RIS) imbalances across elements to enable coherent beamforming and maximize signal quality. In contrast, calibration for positioning and sensing demands higher accuracy, including precise beam response modeling and array geometry calibration, to ensure unbiased estimation of physical parameters such as angle, delay, and location. In~\cite{zhang2023over}, the authors estimate the actual phase shift values of each RIS element under all discrete phase states and address non-ideal phase tuning using an alternating block descent method. RIS phase calibration has also been formulated as an over-the-air phase estimation problem and solved using a backpropagation algorithm~\cite{zhang2024phases}, while efficient calibration and beamforming under discrete RIS phase states were studied in~\cite{borda2025efficient}. In~\cite{zheng2023jrcup}, geometric errors of the RIS such as position and orientation, are calibrated within an RIS-aided positioning system to enhance localization accuracy. However, these studies focus on limited types of impairments or errors, such as phase state and geometry calibration. All hardware impairments and array errors are ultimately reflected in the RIS beam response. Therefore, for high precision positioning, calibration could target the RIS beam response directly rather than isolated hardware parameters. Employing a realistic RIS beam response model can reduce the systematic beam model mismatch. In our previous work~\cite{li2025ris}, three beam models were investigated for RIS-aided positioning to account for different hardware impairments, together with corresponding calibration algorithms. However, multipath effects were not considered, and the validation was limited to a 2D scenario.

Existing studies on array calibration can be broadly categorized into off-line and in-situ (on-site) calibration approaches. Off-line array calibration is based on measurements conducted in controlled environments, such as anechoic chambers~\cite{tang2026large,zhang2024multiprobe,fan2023large,zhang2022phased,li2025phased}. The practical excitations across array elements can be estimated using the rotating element electric field vector (REV) method~\cite{takahashi2012novel,long2016fast} or by solving systems of linear equations~\cite{long2017multi,zhang2019improved}, though these approaches typically require dedicated phase tuning for each element. In~\cite{liu2013unified}, the authors introduce direction-independent correction parameters into the array model to account for mutual coupling effects as well as phase and amplitude mismatches across elements. Offline calibration is generally performed as a standard procedure before commercial deployment. However, the electromagnetic (\ac{em}) characteristics of the array or RIS can change during installation~\cite{ji2021simultaneous}.  

In-situ/on-site array calibration is particularly appealing, as it is performed in the array or RIS’s actual deployment environment rather than under controlled laboratory conditions, thereby capturing the influence of real operating environments. In~\cite{sippel2019situ}, an in-situ calibration method is proposed for a uniform linear array, aiming to correct mutual coupling and geometric errors using an incoherent beacon with narrowband signals. In~\cite{kuric2024technique}, the calibration approach utilizes spherical mode expansion to represent array responses and mitigate platform scattering, relying on a moving antenna array that receives signals from multiple stationary transmitters. In~\cite{pan2023situ}, a direction-dependent array error function is introduced to model various array imperfections for calibration in a positioning system, while multipath components are separated using wideband signals. However, this method is designed for digital antenna arrays, where the received signal at each antenna element or RF chain can be observed, and therefore is not directly applicable to RISs.
     
On-site beam calibration in RIS-aided positioning systems presents additional challenges, since reducing the positioning error floor requires accurate modeling of the RIS beam response and reliable estimation of the corresponding parameters~\cite{wymeersch2020radio,ozturk2022impact}. First, RIS-reflected signals are typically much weaker than the \ac{los} path and can even be comparable in strength to \acp{mpc}~\cite{ellingson2021path}. Accurately extracting the RIS-reflected signals while suppressing \ac{los} and multipath interference is therefore critical but non-trivial. Second, on-site measurements often provide limited flexibility for applying dedicated phase-tuning settings~\cite{wang2021over}. Ideally, the calibration method should work with the RIS default modulations used in its initial operating mode. Third, the method must be scalable to RISs of different sizes, particularly large-scale deployments. More importantly, the calibrated 3D beam model must be accurate enough to reduce the positioning error floor induced by beam model mismatch, rather than improving beamforming gain~\cite{ozturk2022impact,yang2023practical}. These requirements distinguish the considered problem from conventional RIS phase calibration and communication-oriented calibration methods. The existing on-site calibration methods discussed above do not adequately address all these challenges.

\begin{figure}[t]
\centering
\centerline{\includegraphics[width=0.85\linewidth]{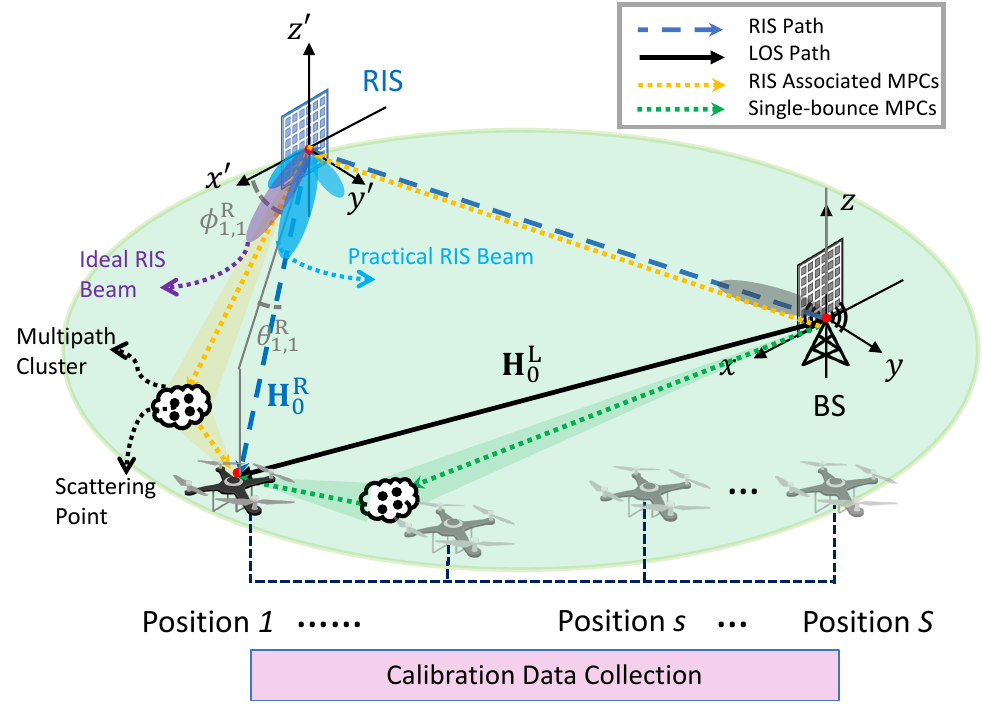}}
\vspace{-0.25cm}
    \caption{An illustration of an RIS-aided positioning system under calibration. The CA is moved to $S$ predefined locations to receive signals, which are then used as inputs for the calibration process.   
}
\label{fig:illustration}
\vspace{-0.2cm}
\end{figure}

In this work, we develop an on-site calibration framework for 3D \ac{ris} beam responses in RIS-aided positioning systems. The main contributions are summarized as follows.

\begin{itemize}
    \item We develop a positioning-oriented on-site \ac{ris} beam calibration model, where the goal is to estimate a practical 3D \ac{ris} beam response model to reduce the positioning error floor induced by beam model mismatch.

    \item We propose an on-site \ac{ris} beam calibration framework that employs a realistic 3D \ac{ris} beam response model accounting for the combined effects of mutual coupling, element patterns, and non-ideal phase tuning networks. The received signals are collected by a calibration agent (CA) at known locations in the deployed environment, where multipath effects are explicitly considered, as illustrated in Fig.~\ref{fig:illustration}.

    \item We develop a two-stage calibration algorithm for estimating the practical \ac{ris} beam response model. The first stage separates the \ac{ris}-reflected channel response from the \ac{los} path and \acp{mpc} through delay-domain sparse recovery. The second stage estimates the required beam model parameters using a gradient-based estimator.

    \item We validate the proposed framework by integrating measured 3D beam patterns from a 16$\times$16 \ac{ris} prototype under 66 phase modulations into simulations. The positioning error floor, characterized by the \ac{alb}, is derived to evaluate the effectiveness of the on-site calibration.
\end{itemize}

    
    
    
The rest of the paper is organized as follows. Section~\ref{sec:signal_model} presents the system model, the practical RIS beam response model, and the problem formulation. Section~\ref{sec:algorithm} details the proposed two-step on-site RIS beam calibration algorithm. Section~\ref{sec:metrics} introduces the evaluation metrics, including the beam response similarity, the positioning performance metric based on the absolute lower bound, and the RIS path channel estimation error metric. Section~\ref{sec:validation} validates the proposed framework using measured 3D RIS beam patterns and numerical simulations. Finally, Section~\ref{sec:conclusion} concludes the paper.

\textit{Notations and Symbols:} 
Italic letters denote scalars (e.g., $a$), bold lower-case letters denote vectors (e.g., $\av$), and bold upper-case letters denote matrices (e.g., $\Am$). The operators $(\cdot)^\top$, $(\cdot)^\herm$, $(\cdot)^*$, $(\cdot)^{-1}$, $\trace(\cdot)$, $\|\cdot\|$, and $\|\cdot\|_\mathrm{F}$ denote the transpose, Hermitian transpose, complex conjugate, inverse, trace, $\ell_2$ norm, and Frobenius norm, respectively. The symbols $\Am \odot \Bm$, $\Am \oslash \Bm$, $\Am \otimes \Bm$, and $\av \circ \bv$ represent the Hadamard product, Hadamard division, Kronecker product, and outer product, respectively. The notation $[\cdot,\ \cdot,\ \cdots,\ \cdot]^\top$ denotes a column vector. The element in the $i$-th row and $j$-th column of a matrix is denoted by $[\cdot]_{i,j}$, while $[\cdot]_{a:b,\,c:d}$ denotes a submatrix formed by rows $a$ to $b$ and columns $c$ to $d$. The operators $\mathrm{Re}\{\cdot\}$ and $\mathrm{Im}\{\cdot\}$ extract the real and imaginary parts of a complex variable, respectively. Finally, $\mathbf{1}_N$ denotes an $N \times 1$ all-ones vector, and $\mathbf{I}_N$ denotes an $N \times N$ identity matrix.
\section{System Model}\label{sec:signal_model}
In this section, we introduce the on-site RIS beam calibration framework, starting with the general system model. Next, we present an RIS beam response model that aims to capture the practical beam behavior by incorporating several key non-ideal factors. Finally, we formulate the positioning task, as the effectiveness of the calibration will ultimately be assessed based on positioning accuracy.

\subsection{System Model}
We consider a three-dimensional (3D) \ac{ofdm} downlink system with one \ac{bs} located at $\pv_B \in \mathbb{R}^3$ and a calibration agent (CA).\footnote{The CA, e.g., a calibrated UE or a drone equipped with measurement instruments, is used only during the calibration phase. Once calibration is completed, the calibrated RIS beam model is used by the positioning estimator for arbitrary UEs.} At the \ac{bs}, we assume a fixed beam directed toward the RIS under calibration. As a result, the BS with an analog antenna array is simplified to a single-antenna model, with its influence reflected in the channel gains of the various propagation paths. The CA is moved across $S$ known locations, $\pv_{C,1}, \ldots, \pv_{C,S} \in \mathbb{R}^3$, to collect pilot reference signals for calibration. The CA positions should be distributed over the angular region of interest. The antenna on the CA is chosen to have a known, fixed radiation pattern, which can be easily calibrated. The RIS, located at $\pv_R \in \mathbb{R}^3$, comprises
$N=N_rN_c$ elements arranged in $N_r$ rows and $N_c$ columns. Its
orientation is parameterized by the Euler angle vector
$\ov_R=[\alpha_R,\beta_R,\gamma_R]^\mathsf{T}$, which defines the
rotation matrix under the extrinsic $x$-$y$-$z$ convention as
\begin{equation}
\Qm_R = \Rm_z(\gamma_R)\Rm_y(\beta_R)\Rm_x(\alpha_R),
\label{eq:ris_rotation_matrix}
\end{equation}
where $\Rm_x(\cdot)$, $\Rm_y(\cdot)$, and $\Rm_z(\cdot)$ denote the
standard right-handed rotation matrices about the $x$-, $y$-, and
$z$-axes, respectively. The \ac{bs} transmits \( G \) \ac{ofdm} symbols over \( K \) subcarriers. In this model, multipath clusters are assumed to be located near
the CA, each comprising several MPCs generated by the
corresponding \acp{sp}. The received signal matrix at the $s$-th CA position, denoted by
$\mathbf{Y}_{s} \in \mathbb{C}^{G \times K}$, is given by
\begin{equation}
    \mathbf{Y}_{s}
    =
    \left(
    \mathbf{H}_{s}^{\mathrm{L}}
    +
    \mathbf{H}_{s}^{\mathrm{SM}}
    +
    \mathbf{H}_{s}^{\mathrm{R}}
    +
    \mathbf{H}_{s}^{\mathrm{RM}}
    \right)
    \odot \mathbf{X}
    +
    \mathbf{N}_{s},
    \label{eq:received_signal}
\end{equation}
where $\mathbf{X} \in \mathbb{C}^{G \times K}$ is the transmitted signal matrix, and
$\mathbf{N}_{s} \in \mathbb{C}^{G \times K}$ denotes the additive white Gaussian noise matrix.

The channel components corresponding to the \ac{los} path, the single-bounce multipath containing $M$ \acp{mpc} (i.e., BS-SP-CA), the \ac{ris}-reflected path, and the \ac{ris}-associated multipath containing $Q$ \acp{mpc} (i.e., BS-SP-RIS-CA or BS-RIS-SP-CA) are modeled as
\begin{align}
    \mathbf{H}_{s}^{\mathrm{L}}
    &=
    \alpha_{s}^{\mathrm{L}}\,
    \mathbf{1}_{G}\,
    \mathbf{d}^{\top}\!\left(\tau_{s}^{\mathrm{L}}\right),
    \label{eq:channel_los}
    \\[4pt]
    \mathbf{H}_{s}^{\mathrm{SM}}
    &=
    \sum_{m=1}^{M}
    \alpha_{s,m}^{\mathrm{SM}}\,
    \mathbf{1}_{G}\,
    \mathbf{d}^{\top}\!\left(\tau_{s,m}^{\mathrm{SM}}\right),
    \label{eq:channel_sm}
    \\[4pt]
    \mathbf{H}_{s}^{\mathrm{R}}
    &=
    \alpha_{s}^{\mathrm{R}}\,
    \mathbf{b}\!\left(\boldsymbol{\varphi}_{s}^{\mathrm{R}}\right)\,
    \mathbf{d}^{\top}\!\left(\tau_{s}^{\mathrm{R}}\right),
    \label{eq:channel_ris}
    \\[4pt]
    \mathbf{H}_{s}^{\mathrm{RM}}
    &=
    \sum_{q=1}^{Q}
    \alpha_{s,q}^{\mathrm{RM}}\,
    \mathbf{b}\!\left(\boldsymbol{\varphi}_{s,q}^{\mathrm{RM}}\right)\,
    \mathbf{d}^{\top}\!\left(\tau_{s,q}^{\mathrm{RM}}\right).
    \label{eq:channel_rm}
\end{align}

Here, $\mathbf{d}(\tau) \in \mathbb{C}^{K \times 1}$ denotes the frequency-domain response vector associated with delay $\tau$, whose $k$-th element is defined as
\begin{equation}
    [\mathbf{d}(\tau)]_{k} = e^{j2\pi k \Delta f \tau},
\end{equation}
where $\Delta f$ is the subcarrier spacing.
The beam response vector $\mathbf{b}(\boldsymbol{\varphi}) \in \mathbb{C}^{G \times 1}$ characterizes the RIS response as a function of the departure angle $\boldsymbol{\varphi}$, where $\boldsymbol{\varphi} = [\phi, \theta]^{\top}$ contains the azimuth and elevation angles. A more detailed discussion of $\boldsymbol{\varphi}$ is provided in the following subsection.
Since the \ac{bs} uses a fixed precoder directed toward the RIS, the \ac{bs} side \ac{aod} steering response and the precoder are absorbed into an effective BS-RIS complex gain, while the RIS side \ac{aoa} steering vector determines the incident field across the RIS elements. In addition, the beam response is assumed to remain constant over the bandwidth of interest.

We consider an asynchronous system in which a clock offset $\tau_{\mathrm{clk}}$ exists between the BS and the CA. Therefore, each delay term includes both the propagation delay and the clock offset. The parameters
$\alpha_{s}^{\mathrm{L}}$ and $\tau_{s}^{\mathrm{L}}$ denote the complex channel gain and delay of the LOS path, respectively;
$\alpha_{s,m}^{\mathrm{SM}}$ and $\tau_{s,m}^{\mathrm{SM}}$ denote those of the $m$-th single-bounce multipath component;
$\alpha_{s}^{\mathrm{R}}$, $\boldsymbol{\varphi}_{s}^{\mathrm{R}}$, and $\tau_{s}^{\mathrm{R}}$ denote the channel gain, departure angle, and delay of the RIS-reflected path;
and $\alpha_{s,q}^{\mathrm{RM}}$, $\boldsymbol{\varphi}_{s,q}^{\mathrm{RM}}$, and $\tau_{s,q}^{\mathrm{RM}}$ denote the corresponding parameters of the $q$-th RIS-associated multipath component.
\subsection{Beam Response Model}
We first present the simplified beam model of an \ac{ris} or antenna array commonly used in previous studies\cite{wu2024employing,zhang2023approximate,keykhosravi2022ris}. This model does not account for hardware impairments or array errors, and assumes isotropic element patterns. It serves as a baseline for evaluating positioning performance in the absence of calibration. The model is defined as

\begin{equation}\label{eq:b_ideal}
    \tilde{\bv}(\varphiv) = \tilde{\Wm}^\herm \av(\varphiv),
\end{equation}
where \( \av(\varphiv) \in \mathbb{C}^{N \times 1} \) denotes the steering vector, defined as
\begin{equation}
[\av(\phi,\theta)]_n
=
\exp\!\left(j\frac{2\pi}{\lambda}\left(n_c d_c \sin\theta\cos\phi+n_r d_r \cos\theta\right)\right),
\end{equation}
with the \( n \)-th RIS element located at the \( n_c \)-th column and \( n_r \)-th row. Here, \( d_c \) and \( d_r \) represent the spacing of the elements along the columns and rows, respectively and $\lambda$ is the free-space wavelength at center frequency.
The steering vector characterizes the phase difference due to impinging or reflected waves as influenced by the RIS configuration. The codebook matrix \( \tilde{\Wm} = [\tilde{\wv}_1, \tilde{\wv}_2, \ldots, \tilde{\wv}_G] \in \mathbb{C}^{N \times G} \) contains the predefined phase tuning values applied to the RIS elements across different symbols.

To more accurately represent the true RIS beam response, we extend the beam model proposed in equation (3) of our previous work~\cite{li2025ris} to a full 3D pattern representation. The model presented here is designed to capture the combined effects of various hardware impairments and is formulated as
\begin{equation}
    \bv( \varphiv) = \Wm^\herm \left( \av(\varphiv) \odot \gv( \varphiv) \right),
    \label{eq:b_proposed}
\end{equation}
where \( \gv(\varphiv) \in \mathbb{C}^{N \times 1} \), with its \( n \)-th element denoted by \( [\gv(\varphiv)]_n \), represents the realistic complex element pattern of the \( n \)-th RIS element. The matrix \( \Wm \in \mathbb{C}^{N \times G} \) contains the actual phase tuning values applied to the RIS elements, incorporating the combined effects of mutual coupling and non-ideal phase tuning networks. 
For large-scale RIS configurations, the element pattern vector $\boldsymbol{g}(\boldsymbol{\varphi})$ can be approximated by a scalar function $g(\boldsymbol{\varphi})$, as most elements exhibit similar patterns due to nearly identical \ac{em} boundary conditions~\cite{singh2013mutual}. Consequently, the beam model in~\eqref{eq:b_proposed} neglecting the edge elements effects can be simplified as
\begin{equation}
    \mathbf{b}(\boldsymbol{\varphi}) = g(\boldsymbol{\varphi})\, \mathbf{W}^{\herm
    } \mathbf{a}(\boldsymbol{\varphi}).
    \label{eq:b_proposed2}
\end{equation}
We use the model in~\eqref{eq:b_proposed2} as the generative
beam model in the numerical results, since RISs are typically deployed at large scales to ensure sufficient reflected signal strength~\cite{bjornson2019intelligent}.

Considering all \( S \) CA positions used for calibration, the beam response vectors can be stacked to form the beam response matrix \( \Bm \in \mathbb{C}^{G \times S} \) as
\begin{equation}
     \Bm = \left[ \bv(\varphiv^{\mathrm{R}}_{1}), \ldots, \bv(\varphiv^{\mathrm{R}}_{S}) \right].
    \label{eq:cali_model_1}
\end{equation}
However, in practical positioning or calibration processes, the RIS path channel gain and the element pattern response cannot be estimated separately. For notational convenience and to facilitate the presentation of the calibration algorithm, we incorporate the RIS-reflected path gain into the beam response model and rewrite the beam response matrix as
\begin{equation}
     \Bm = \Wm^\herm \Am(\Phim)\Gammam,
    \label{eq:cali_model_2}
\end{equation}
where \( \Am(\Phim) = \left[ \av(\varphiv^{\mathrm{R}}_{1}), \ldots, \av(\varphiv^{\mathrm{R}}_{S}) \right] \in \mathbb{C}^{N \times S} \) denotes the steering matrix, with \( \Phim = \left[ \varphiv^{\mathrm{R}}_{1}, \ldots, \varphiv^{\mathrm{R}}_{S} \right] \). The diagonal matrix \( \Gammam \in \mathbb{C}^{S \times S} \) captures the combined effects of the RIS-path channel gain and the element pattern response, and its \( s \)-th diagonal element is given by
\[
[\Gammam]_{s,s} = {\gamma}_s = \alpha_{s}^{\mathrm{R}} \, g(\varphiv^{\mathrm{R}}_{s}).
\]

\subsection{Problem Statement}
Given the measurements collected at the known CA locations, the objective of on-site RIS beam calibration is to estimate the parameters of the 3D RIS beam response model in~\eqref{eq:cali_model_2}. Specifically, the calibration problem is to estimate the RIS modulation matrix $\Wm$ and the diagonal gain-pattern matrix $\Gammam$ from $\{\mathbf{Y}_s\}_{s=1}^{S}$, while accounting for the coexistence of the RIS-reflected path, the \ac{los} path, and \acp{mpc}. The calibrated model is then used by the positioning estimator and should be sufficiently accurate over the angular region of interest, since residual beam-model mismatch may induce a positioning error floor.
\section{Calibration Algorithm}\label{sec:algorithm}
This section presents the proposed on-site RIS beam calibration algorithm, which estimates the required parameters in \eqref{eq:b_proposed}. As highlighted in Section~\ref{sec:intro}, a major challenge in on-site RIS beam calibration is the separation of the RIS-reflected signal from the received measurements at the CA, while mitigating interference from the \ac{los} path and other multipath components. Direct application of conventional sparse recovery methods often fails in this setting due to the lack of structural exploitation.
To address this, we propose a structure-aware delay-domain extraction method that combines (i) noncoherent aggregation across RIS configurations, (ii) high-resolution delay refinement, and (iii) geometry-assisted path identification. The unknown parameters of the RIS beam response model are then estimated using a gradient descent-based optimization approach. The details of each step are presented below.

Before presenting the calibration algorithm, we define the path gain vector
$\boldsymbol{\xi}_s \in \mathbb{C}^{G}$ at the $s$-th CA location. For the \ac{los} path and non-RIS associated \acp{mpc}, $\boldsymbol{\xi}_s$ contains the corresponding complex channel gains observed under the $G$ codewords. For RIS-related paths, the entries of $\boldsymbol{\xi}_s$ represent effective complex gains that include both the corresponding channel gain and the RIS beam response induced by each codeword. In particular,
$\boldsymbol{\xi}^{\mathrm{R}}_s$ denotes the RIS-reflected path gain vector at the $s$-th CA location and corresponds to the $s$-th column of
$\Bm$.
\subsection{Step 1: Geometry-Aware RIS Path Extraction via Delay-Domain Sparse Recovery}
\label{subsec:step1_omp}
In addition to covering the angular range of interest, the number of CA samples \(S\) must be large enough to make the calibration problem observable. From \eqref{eq:cali_model_2}, the measurement matrix provides \(GS\) complex observations, while the unknowns consist of \(GN\) entries in \(\Wm\) and \(S\) diagonal entries in \(\Gammam\). Therefore, a necessary counting condition for estimating all unknowns is
\begin{equation}
    GS \geq GN + S
    \quad \Longleftrightarrow \quad
    S \geq \frac{GN}{G-1}, \qquad G>1.
    \label{eq:S_condition}
\end{equation}
The impact of the selected number of CA samples \(S\) is investigated later in Section~\ref{sec:validation}-B. In this work, the CA locations are placed on a uniform angular grid in the azimuth and elevation domains, since this provides a general sampling strategy that is compatible with arbitrary beamforming codebooks. 
For the $s$-th calibration position, the pilot symbols are first removed by Hadamard division:
\begin{equation}
   \tilde{\Ym}_s^{(0)} = \Ym_s \oslash \Xm,
   \label{eq:pilot_remove}
\end{equation}
where $\tilde{\Ym}_s^{(0)} \in \mathbb{C}^{G\times K}$ denotes the pilot-removed frequency-domain observation. Let $\tilde{\Ym}_s^{(i)}$ denote the residual observation after the $i$-th iteration.
\subsubsection{Noncoherent Multi-Codeword Aggregation}
Since the RIS response varies across codewords while the propagation delay remains invariant for a given CA position, we enhance path detectability by noncoherently aggregating the delay-domain power spectra over all codewords. For the $g$-th codeword at the $s$-th CA position, the delay domain residual response at iteration $i$ is obtained by applying a $K$-point FFT across the subcarriers. Its $n$-th delay bin sample is
\begin{equation}
    p_{s,g}^{(i)}[n]
    =
    \sum_{k=0}^{K-1}
    [\tilde{\Ym}_s^{(i)}]_{g,k}
    e^{-j2\pi kn/K},
    \qquad n=0,\ldots,K-1.
    \label{eq:delay_bin_fft}
\end{equation}
The corresponding non-coherent power delay profile is
\begin{equation}
    P_{s,i}[n]
    =
    \sum_{g=1}^{G}
    \left|
    p_{s,g}^{(i)}[n]
    \right|^2,
    \qquad n=0,\ldots,K-1,
    \label{eq:pdp_noncoherent}
\end{equation}
where the $n$-th delay bin corresponds to $\tau_n = n/(K\Delta f)$, with $\Delta f$ denoting the subcarrier spacing.

A coarse estimate of the $i$-th path delay at the $s$-th CA position is obtained as
\begin{equation}
   \tau_{s,i}^{\mathrm{c}} = \arg\max_{\tau} P_{s,i}(\tau).
   \label{eq:coarse_delay_new}
\end{equation}

\subsubsection{High-Resolution Delay Refinement}
To refine this estimate, a local dense delay grid centered around $\tau_{s,i}^{\mathrm{c}}$ is defined as
\begin{equation}
   \mathcal{T}_{s,i}=\{\tilde{\tau}_{s,i,j}\}_{j=1}^{N_{\tau}},
\end{equation}
with endpoints
\begin{equation}
   \tilde{\tau}_{s,i,1} = \tau_{s,i}^{\mathrm{c}} - \Delta \tau,
   \qquad
   \tilde{\tau}_{s,i,N_{\tau}} = \tau_{s,i}^{\mathrm{c}} + \Delta \tau,
   \label{eq:delay_grid_new}
\end{equation}
where \(\Delta \tau = 1/\mathrm{BW}\), and \(\mathrm{BW}\) denotes the total \ac{ofdm} signal bandwidth. The refined delay estimate is selected by
\begin{equation}
   \hat{\tau}_{s,i}
   =
   \arg\max_{\tau \in \mathcal{T}_{s,i}}
   \sum_{g=1}^{G}
   \left|
   \dv(\tau)^{\herm}
   [\tilde{\Ym}_s^{(i)}]_{g,:}^{\top}
   \right|^2.
   \label{eq:refined_delay_new}
\end{equation}

\subsubsection{Residual Update, Stopping Criterion, and Geometry-Assisted Path Identification}
After the \(i\)-th path delay \(\hat{\tau}_{s,i}\) is detected, the coefficients of all detected paths are re-estimated jointly as
\begin{equation}
    \{\hat{\xiv}_{s,\ell}\}_{\ell=0}^{i}
    =
    \arg\min_{\{\xiv_{s,\ell}\}_{\ell=0}^{i}}
    \left\|
    \tilde{\Ym}_s^{(0)}
    -
    \sum_{\ell=0}^{i}
    \xiv_{s,\ell}
    \dv(\hat{\tau}_{s,\ell})^{\top}
    \right\|_{\mathrm{F}}^{2}.
    \label{eq:ls_xi_new}
\end{equation}
The residual is updated as
\begin{equation}
    \tilde{\Ym}_s^{(i+1)}
    =
    \tilde{\Ym}_s^{(0)}
    -
    \sum_{\ell=0}^{i}
    \hat{\xiv}_{s,\ell}
    \dv(\hat{\tau}_{s,\ell})^{\top}.
    \label{eq:residual_update}
\end{equation}
The iterations stop when
\begin{equation}
   i+1 \geq I_{\max}
    \quad \text{or} \quad
    \frac{\|\tilde{\Ym}_s^{(i+1)}\|_{\mathrm{F}}^2}
    {\|\tilde{\Ym}_s^{(0)}\|_{\mathrm{F}}^2}
    \leq \epsilon_{\mathrm{res}}.
    \label{eq:stopping_criterion}
\end{equation}
The resulting number of detected paths is denoted by \(I_s\), and
\begin{equation}
    \tilde{\Ym}_s^{(0)}
    =
    \sum_{i=0}^{I_s-1}
    \hat{\xiv}_{s,i}
    \dv(\hat{\tau}_{s,i})^{\top}
    +
    \tilde{\Ym}_s^{(I_s)}.
    \label{eq:channel_decomp_new}
\end{equation}

With known BS, RIS, and CA locations, the LOS and RIS-reflected propagation distances at the \(s\)-th CA position are denoted by \(d_{s,\mathrm{LOS}}\) and \(d_{s,\mathrm{RIS}}\), respectively. Assuming the LOS path is present and detected, the earliest detected path is used as the LOS reference. Since the clock offset is common to all paths at the same CA position, it is removed by taking delay differences. The distance difference of the \(i\)-th detected path is
\begin{equation}
    \Delta d_{s,i} = c\big(\hat{\tau}_{s,i} - \hat{\tau}_{s,\mathrm{LOS}}\big),
    \qquad
    \delta_{s,\mathrm{ref}} = d_{s,\mathrm{RIS}} - d_{s,\mathrm{LOS}}.
    \label{eq:distance}
\end{equation}
The candidate RIS-reflected paths are selected as
\begin{equation}
    \mathcal{I}_s^{\mathrm{R}}
    =
    \left\{
    i\in\{0,\ldots,I_s-1\}:
    \left| \Delta d_{s,i} - \delta_{s,\mathrm{ref}} \right| \le \varepsilon
    \right\}.
\end{equation}
The RIS-reflected path index is then
\begin{equation}
    i_s^\star
    =
    \arg\min_{i\in\mathcal{I}_s^{\mathrm{R}}}
    \left| \Delta d_{s,i} - \delta_{s,\mathrm{ref}} \right|.
\end{equation}
Finally,
\begin{equation}
    \hat{\tau}_s^{\mathrm{R}} = \hat{\tau}_{s,i_s^\star},
    \qquad
    \hat{\xiv}_s^{\mathrm{R}} = \hat{\xiv}_{s,i_s^\star}.
\end{equation}
The estimated RIS beam response matrix used in the second calibration step is formed as
\begin{equation}
    \hat{\Bm}
    =
    \big[
    \hat{\xiv}_1^{\mathrm{R}},
    \ldots,
    \hat{\xiv}_S^{\mathrm{R}}
    \big]
    \in \mathbb{C}^{G\times S}.
    \label{eq:Bhat_step1_new}
\end{equation}

\subsection{Step 2: Gradient-Based Estimation of RIS Beam Model Parameters}
\label{subsec:step2_gd}

In the second step of the proposed calibration framework, the unknown parameters of the RIS beam response model in \eqref{eq:b_proposed2} are estimated using the RIS-path responses extracted in Step~1. The goal is to jointly estimate the effective RIS codebook and the combined complex gains by fitting the model to the measured beam responses.

\subsubsection{Calibration Problem Formulation}

Recall from \eqref{eq:cali_model_2} that the estimated RIS beam response matrix satisfies
\begin{equation}
    \hat{\Bm} \approx \Wm^\herm \Am(\Phim)\Gammam,
    \label{eq:cali_model_est}
\end{equation}
where \(\hat{\Bm}\in\mathbb{C}^{G\times S}\) is obtained from Step~1, \(\Wm\in\mathbb{C}^{N\times G}\).
The calibration problem is formulated as
\begin{equation}
    \mathcal{L}(\Wm,\Gammam)
    =
    \left\|
    \hat{\Bm} - \Wm^\herm \Am(\Phim)\Gammam
    \right\|_\mathrm{F}^{2},
    \label{eq:loss_step2}
\end{equation}
and the corresponding optimization problem is
\begin{equation}
    [\hat{\Wm},\hat{\Gammam}]
    =
    \arg\min_{\Wm,\Gammam}
    \mathcal{L}(\Wm,\Gammam)
    \quad
    \text{s.t. }
    \|\wv_g\|_2=1,\ \forall g,
    \label{eq:opt_step2}
\end{equation}
where \(\wv_g\) denotes the \(g\)-th column of \(\Wm\). The unit norm constraint\footnote{This normalization removes the scale ambiguity between \(\Wm\) and \(\Gammam\) in the factorization \(\Bm=\Wm^\herm\Am(\Phim)\Gammam\). It can also be viewed as a bounded power approximation for passive RIS codewords.}
 is imposed to obtain an identifiable factorization.

The optimization is initialized by setting \(\Wm^{(0)}=\tilde{\Wm}\), where \(\tilde{\Wm}\) denotes the ideal RIS codebook, and by setting \(\Gammam^{(0)}=\mathbf{I}_S\). This initialization provides a physically meaningful starting point and improves convergence.

\subsubsection{Gradient-Based Optimization}

Since \eqref{eq:opt_step2} is not jointly convex in \(\Wm\) and \(\Gammam\), we adopt an alternating optimization strategy. Specifically, given the current estimate of \(\Gammam\), the matrix \(\Wm\) is updated by gradient descent; then, given the updated \(\Wm\), the diagonal entries of \(\Gammam\) are updated in closed form.

For notational convenience, let
\begin{equation}
    \av_s \triangleq [\Am(\Phim)]_{:,s}\in\mathbb{C}^{N\times 1},
    \qquad
    \hat{\bv}_s \triangleq [\hat{\Bm}]_{:,s}\in\mathbb{C}^{G\times 1},
\end{equation}
for \(s=1,\ldots,S\). Then \eqref{eq:cali_model_est} can be written column-wise as
\begin{equation}
    \hat{\bv}_s \approx \gamma_s\, \Wm^\herm \av_s.
    \label{eq:column_model}
\end{equation}

\paragraph{Gradient Update of \(\Wm\)}

Given \(\Gammam^{(i)}\), the matrix \(\Wm\) is updated by
\begin{equation}
    \Wm^{(i+1)}
    =
    \Wm^{(i)}
    -
    l_r
    \left.
    \frac{\partial \mathcal{L}}{\partial \Wm^*}
    \right|_{\Wm=\Wm^{(i)},\,\Gammam=\Gammam^{(i)}},
    \label{eq:W_update}
\end{equation}
where \(l_r>0\) is the learning rate. Using Wirtinger calculus, the gradient is given by
\begin{equation}
    \frac{\partial \mathcal{L}}{\partial \Wm^*}
    =
    \sum_{s=1}^{S}
    \left(
    -\gamma_s\, \av_s \hat{\bv}_s^\herm
    +
    |\gamma_s|^2 \av_s \av_s^\herm \Wm
    \right).
    \label{eq:gradient_W}
\end{equation}
Optionally, the summation in \eqref{eq:gradient_W} may be scaled by \(1/S\) without changing the minimizer.

After each gradient step, the columns of \(\Wm\) are projected onto the unit sphere:
\begin{equation}
    \wv_g^{(i+1)}
    \leftarrow
    \frac{\wv_g^{(i+1)}}{\|\wv_g^{(i+1)}\|_2},
    \qquad g=1,\ldots,G.
    \label{eq:W_projection}
\end{equation}

\paragraph{Closed-Form Update of \(\Gammam\)}

Given \(\Wm^{(i+1)}\), each diagonal coefficient \(\gamma_s\) is updated independently by solving
\begin{equation}
    \gamma_s^{(i+1)}
    =
    \arg\min_{\gamma_s\in\mathbb{C}}
    \left\|
    \hat{\bv}_s - \gamma_s \Wm^{(i+1)\herm}\av_s
    \right\|_2^2.
    \label{eq:gamma_ls}
\end{equation}
The least-squares solution is
\begin{equation}
    \gamma_s^{(i+1)}
    =
    \frac{
    \left(\Wm^{(i+1)\herm}\av_s\right)^\herm \hat{\bv}_s
    }{
    \left\|\Wm^{(i+1)\herm}\av_s\right\|_2^2 + \epsilon_0
    },
    \label{eq:gamma_update}
\end{equation}
where \(\epsilon_0>0\) is a small regularization constant introduced to improve numerical stability. The alternating updates in \eqref{eq:W_update}--\eqref{eq:gamma_update} are repeated until either a maximum number of epochs is reached or the relative decrease of the loss in \eqref{eq:loss_step2} falls below a prescribed threshold. The final estimates are denoted by \(\hat{\Wm}\) and \(\hat{\Gammam}\).

The above calibration procedure is scalable to large RIS dimensions and three-dimensional angular domains, while effectively capturing the combined effects of mutual coupling, element pattern distortion, and non-ideal phase tuning.
\subsection{Computational Complexity}
Let \(I_{\max}\) denote the maximum number of detected paths for each CA position,
\(N_{\tau}\) the number of points in the local dense delay grid, and \(N_{\mathrm{ep}}\)
the maximum number of epochs used in Step~2. In Step~1, for each CA position and
each iteration, the delay-domain transformation over \(G\) codewords requires
\(\mathcal{O}(GK\log K)\) operations, while the high resolution delay refinement
requires \(\mathcal{O}(GN_{\tau}K)\) operations. The joint least-squares update
for the detected path coefficients has complexity
\(\mathcal{O}(K i^2+i^3+GKi)\) at the \(i\)-th iteration, where the delay
dictionary is shared across all codewords. Here, the term \(K i^2\) corresponds
to forming the Gram matrix of the \(i\) detected delay atoms, \(i^3\) corresponds
to solving the resulting \(i\times i\) least-squares system, and \(GKi\)
corresponds to projecting the \(G\) codeword observations onto the detected
delay subspace. Therefore, the overall complexity of
Step~1 over all \(S\) CA positions is
\begin{equation}
\mathcal{O}\!\left(
S \sum_{i=1}^{I_{\max}}
\left(
GK\log K
+
GN_{\tau}K
+
K i^2
+
i^3
+
GKi
\right)
\right).
\end{equation}
Since \(I_{\max}\) is typically small, the dominant terms are the delay-domain
search and refinement, yielding approximately
\begin{equation}
\mathcal{O}\!\left(
S I_{\max} G K(\log K+N_{\tau})
\right).
\end{equation}

In Step~2, the steering matrix \(\Am(\Phim)\in\mathbb{C}^{N\times S}\) can be
precomputed with complexity \(\mathcal{O}(NS)\). For each epoch, evaluating the
gradient with respect to \(\Wm\) requires \(\mathcal{O}(SNG)\) operations, and
the closed-form update of the \(S\) diagonal entries of \(\Gammam\) also requires
\(\mathcal{O}(SNG)\) operations.  Since the one-time precomputation cost is
dominated by the iterative updates, the overall complexity of Step~2 is
\begin{equation}
\mathcal{O}\!\left(
N_{\mathrm{ep}}SNG
\right).
\end{equation}
Overall, the proposed calibration algorithm scales linearly with the number of
CA samples \(S\), the number of codewords \(G\), and the RIS size \(N\) in the
model-parameter estimation step, which makes it suitable for large scale RIS
calibration.
\section{Evaluation Metrics}\label{sec:metrics}
This section defines the metrics used to evaluate the proposed calibration framework. The beam response similarity (\ac{brs}) assesses the agreement between the estimated and true RIS beam responses, thereby quantifying the fidelity of the calibrated RIS beam model. The positioning metric evaluates the resulting localization accuracy after calibration. In addition, we define a diagnostic channel-estimation error metric for reflected RIS path in Step~1 of the calibration algorithm, since these quantities serve as important intermediate inputs to the beam calibration step.

\subsection{Beam Response Similarity}
Intuitively, a calibrated RIS beam that closely matches the true RIS beam indicates reduced RIS beam model mismatch and validates the fidelity of the employed RIS beam model.
In the field of over-the-air testing, the power angular spectrum similarity percentage is widely adopted as a standard metric for evaluating the similarity of angular power distributions between a reference and a reconstructed channel~\cite{3gppTR38827}. Motivated by this metric, we extend the concept to the complex-valued beam response to assess the similarity between the ground-truth and calibrated RIS beam. The vector valued RIS beam response in~\eqref{eq:b_proposed2} collects the responses over all RIS modulation patterns. For a given codeword, we denote the scalar beam responses of the estimated and true RIS beams over the entire solid-angle domain by $\hat b(\varphiv)$ and $\bar b(\varphiv)$, respectively. The beam response similarity (\ac{brs}) is then defined as
\begin{equation}
    M_\mathrm{BRS} = 1-\frac{1}{2}\int_{\boldsymbol{\varphi}} 
    \left|
    \frac{\hat b(\varphiv)}{\int_{\boldsymbol{\varphi}}|\hat b(\varphiv)| \text{d}\varphiv} - 
    \frac{\bar b(\varphiv)}{\int_{\boldsymbol{\varphi}}|\bar b(\varphiv)| \text{d}\varphiv}
    \right|
    \text{d}\varphiv,
    \label{eq:M_brs_1}
\end{equation}
where $\hat b(\varphiv)$ and $\bar b(\varphiv)$ denote the beam response function under evaluation and the ground-truth beam response function, respectively. Both functions are normalized by their corresponding integrals in \eqref{eq:M_brs_1} before computing the similarity.
However, in practice, the CA can only be moved to a finite number of locations. To improve measurement efficiency, we restrict the evaluation to the angular range of interest and use a total of \(S\) sampled directions for calibration. We define the following finite-sample BRS metric as
\begin{equation}
    M_{\mathrm{BRS},g}
    =
    1-\frac{1}{2}
    \left\|
    \frac{[\hat{\Bm}]_{g,:}}{\|[\hat{\Bm}]_{g,:}\|_\mathrm{F}}
    -
    \frac{[\bar{\Bm}]_{g,:}}{\|[\bar{\Bm}]_{g,:}\|_\mathrm{F}}
    \right\|_\mathrm{F}.
    \label{eq:M_brs_2}
\end{equation}
Here, \( [\hat{\Bm}]_{g,:} \) and \( [\bar{\Bm}]_{g,:} \) denote the \(g\)-th rows of \( \hat{\Bm} \) and \( \bar{\Bm} \) according to \eqref{eq:cali_model_2}, respectively, i.e., the beam responses of the \(g\)-th codeword over the \(S\) sampled directions. The metric \( M_{\mathrm{BRS},g} \in [0,1] \) measures the similarity between the estimated and ground-truth beam responses for the \(g\)-th codeword, where 0 indicates completely dissimilar responses and 1 indicates perfect agreement.
The overall calibration performance across the codebook is quantified by the average BRS, defined as
\begin{equation}
    \bar{M}_{\mathrm{BRS}}
    =
    \frac{1}{G}\sum_{g=1}^{G} M_{\mathrm{BRS},g},
    \label{eq:M_brs_avg}
\end{equation}
where \(G\) denotes the total number of codewords.
\subsection{Positioning Performance Metric}
After calibration, the RIS beam response model can be updated in the positioning estimator, thereby reducing the expected model mismatch. The mismatched lower bound (MLB) can quantify the impact of using a mismatched beam response model on the position estimation. Assume a UE is located at \( \pv_U \in \mathbb{R}^3 \). The received signals $\Ym \in \mathbb{C}^{G \times K}$ at the UE can also be derived using equation~\eqref{eq:received_signal}-\eqref{eq:channel_ris} by replacing CA with UE. We define the channel parameter vector as $\boldsymbol{\eta} = [\tau_\mathrm{L}, \rho_\mathrm{L}, \beta_\mathrm{L}, \phi_\mathrm{R}, \theta_\mathrm{R}, \tau_\mathrm{R}, \rho_\mathrm{R}, \beta_\mathrm{R}]^\top$, which contains the channel parameters associated with the LOS and RIS paths. In the vector $\boldsymbol{\eta}$, $\rho$ and $\beta$ denote the amplitude and phase of the complex channel gain, respectively, i.e., $\alpha = \rho e^{-j\beta}$. We denote $\mathbf{y}=\operatorname{vec}(\mathbf{Y})\in \mathbb{C}^{GK}$ 
as the received signal vector.\footnote{The multipath is neglected in the mismatch analysis as we focus on the impacts caused by beam response model mismatch.} 

The pseudo-true parameter can be interpreted as the best parameter value attainable within the assumed RIS beam model. When the model is correctly specified, it coincides with the true parameter. Otherwise, it captures the closest model-based approximation to the true RIS beam response. The remaining mismatch produces a deterministic estimation bias, which leads to a non-zero positioning error floor.
 We first derive the pseudo-true parameter vector for channel parameter $\etav$ as
\begin{align}
    {\boldsymbol\eta}_0 = \arg \min_{\boldsymbol\eta} D_\text{KL}(f_\text{TM}(\yv|\bar {\boldsymbol\eta})\Vert f_\text{MM}(\yv| {\boldsymbol\eta})),
\end{align}
where \( f_{\text{TM}}( \yv \mid \bar{\etav}) \) and \( f_{\text{MM}}( \yv \mid \etav) \) denote the probability density functions (PDFs) of the true model (using the true beam response \( \bar{\bv} \)) and the mismatched model (using either the calibrated beam response \( \bv \) or the simplified beam model \( \bv_0 \)), respectively. Here, \( \bar{\boldsymbol{\eta}} \) represents the vector of true channel parameters. 
As the pseudo-true parameter vector is the parameter value within the assumed model that best approximates the true distribution, it can be obtained as
\begin{equation}
    {\boldsymbol\eta}_0
    = \arg \min_{{\boldsymbol\eta}} \Vert \epsilonv(\etav) \Vert^2 
    = \arg \min_{{\boldsymbol\eta}} \Vert \bar\muv(\bar {\boldsymbol\eta}) - \muv({\boldsymbol\eta})\Vert^2
    \label{eq:pseudotrue_ch},
\end{equation}
where $\bar\muv(\bar {\boldsymbol\eta})$ and $\muv({\boldsymbol\eta})$ are the noise-free received signal vectors using the true beam response and the employed beam model, respectively. The state vector is defined as $\mathbf{s} = [\mathbf{p}_{U}^\top, \tau_{\mathrm{clk}}]^\top$, which includes the 3D position of the UE and the clock offset. The MLB for the state vector is given by
\begin{equation}
    \mathrm{MLB}_{\mathbf{s}}(\mathbf{s}_0)
    =
    \mathrm{MCRB}(\mathbf{s}_0)
    +
    \underbrace{(\bar{\mathbf{s}}-\mathbf{s}_0)(\bar{\mathbf{s}}-\mathbf{s}_0)^\top}_{\text{bias term}}
\end{equation}
where $\bar{\mathbf{s}}$ and $\mathbf{s}_0$ are the true and pseudo-true state parameter vectors, respectively. The pseudo-true state vector $\mathbf{s}_0$ can be obtained as~\cite{chen2023modeling}
\begin{equation}
\mathbf{s}_0 =
\arg\min_{\mathbf{s}}
\left( \boldsymbol{\eta}_{0} - \boldsymbol{\eta}(\mathbf{s}) \right)^\top
\boldsymbol{\mathcal{I}}(\bar{\boldsymbol{\eta}})
\left( \boldsymbol{\eta}_{0} - \boldsymbol{\eta}(\mathbf{s}) \right),
\end{equation}
where
\begin{equation}
\mathcal{I}(\etav)
=
\frac{2}{\sigma_n^2}
\sum_{g=1}^{G}
\sum_{k=1}^{K}
\operatorname{Re}
\left\{
\left(
\frac{\partial \mu_{g,k}}{\partial \etav}
\right)^{\mathrm{H}}
\left(
\frac{\partial \mu_{g,k}}{\partial \etav}
\right)
\right\}.
\label{eq:fim_eta}
\end{equation}
Here, $\mathcal{I}(\etav) \in \mathbb{R}^{8\times8}$ is the Fisher information matrix~\cite{kay1993statistical} (Sec. 3) of the channel parameter estimation. \( \bar{\sv} \) denotes the true state vector. Gradient-based estimation methods can be used as a practical tool to estimate the $\etav_0$. 

The MLB consists of two terms: the MCRB term and the bias term. The MCRB represents the variance-related component and varies with SNR similarly to the conventional \ac{crb}. In contrast, the bias term is independent of transmit power and captures the systematic error caused by model mismatch. Since this work focuses on the positioning error floor induced by beam response model mismatch, we define the \ac{alb} using the UE position component of the bias term. Let $\mathbf{A}_{s}=[\mathbf{I}_{3},\mathbf{0}_{3\times 1}]$ denote the selection matrix that extracts the UE position from the state vector. The \ac{alb} is then defined as
\begin{equation}
\mathrm{ALB}
=
\sqrt{\mathrm{tr}\!\left(
\mathbf{A}_{s}
(\bar{\mathbf{s}}-\mathbf{s}_0)(\bar{\mathbf{s}}-\mathbf{s}_0)^\top
\mathbf{A}_{s}^{\top}
\right)}.
\end{equation}
\subsection{Channel Estimation Error}
To diagnose the intermediate output of Step~1 of the calibration algorithm, we also evaluate the channel estimation error for the reflected RIS path, which contains the useful information for calibration. Let \(\bar{\Bm}\) and \(\hat{\Bm}\) denote the true and estimated RIS-related effective beam matrices, respectively. The entry \([\bar{\Bm}]_{g,s}\) corresponds to the RIS-related effective path gain observed at the \(s\)-th CA location under the \(g\)-th codeword, while \([\hat{\Bm}]_{g,s}\) denotes its estimate. The squared error for the \(g\)-th codeword and the \(s\)-th CA location is defined as
\begin{equation}
    e_{g,s}^{\mathrm{R}}
    =
    \left|[\hat{\Bm}]_{g,s}-[\bar{\Bm}]_{g,s}\right|^2.
    \label{eq:channel_squared_error}
\end{equation}
This metric is used only to assess the accuracy of the RIS path channel estimation in Step~1, whereas the final calibration performance is evaluated using the beam response and positioning metrics.

\section{Validation and Discussions}\label{sec:validation}
This section validates the proposed calibration framework in two steps. First, we present the RIS beam pattern measurements conducted in an anechoic chamber and analyze their deviations from the ideal model to motivate the need for calibration. Second, these measurements are incorporated into a simulation framework to evaluate how calibration improves beam reconstruction and reduces the positioning error floor.

\subsection{3D RIS Beam Patterns}
\begin{figure}[!t]
\centering
\centerline{\includegraphics[width=1\linewidth]{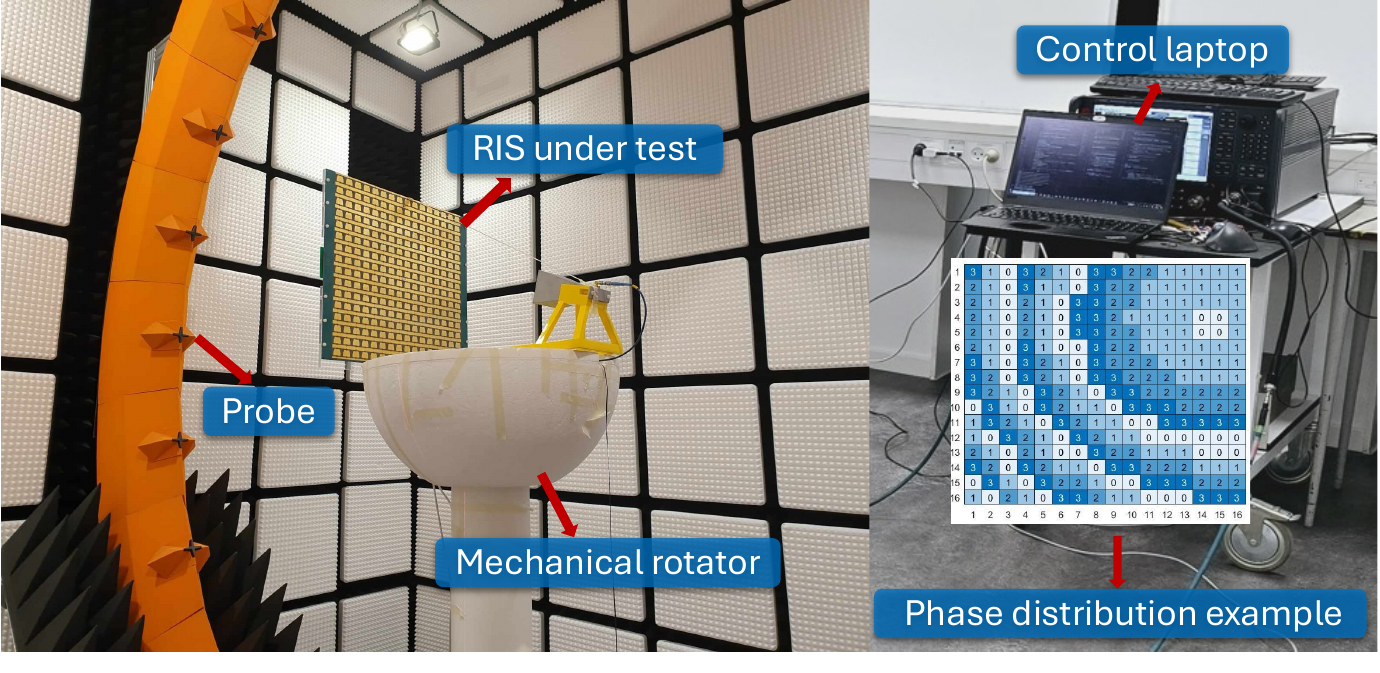}}
\vspace{-0.1cm}
\caption{3D RIS beam pattern measurement setup, showing the setup in the anechoic chamber on the left and the control laptop used for phase tuning on the right. The entries 0, 1, 2, and 3 denote the four RIS phase tuning states, corresponding to $0^\circ$, $90^\circ$, $180^\circ$, and $270^\circ$, respectively.}
\label{fig:RISmea_setup}
\vspace{-0.2cm}
\end{figure}

\begin{table}[t]
\centering
\caption{3D RIS Beam Pattern Measurement Configuration}
\label{tab:measurement_settings}
\footnotesize
\setlength{\tabcolsep}{4pt}
\renewcommand{\arraystretch}{1.12}
\begin{tabular}{|l|c|}
\hline
\textbf{Parameter} & \textbf{Value} \\ \hline
Center frequency & \(5~\mathrm{GHz}\) \\ \hline
Preset steering range, azimuth & \([-50^\circ,50^\circ]\) \\ \hline
Preset steering range, elevation & \([-50^\circ,0^\circ]\) \\ \hline
Preset steering step & \(10^\circ\) \\ \hline
Number of preset steering directions & \(66\) \\ \hline
Angular sampling resolution & \(1^\circ\) \\ \hline
\end{tabular}
\end{table}

\begin{figure*}[t]
\centering

\begin{minipage}{0.32\linewidth}
\centering
\includegraphics[width=\linewidth]{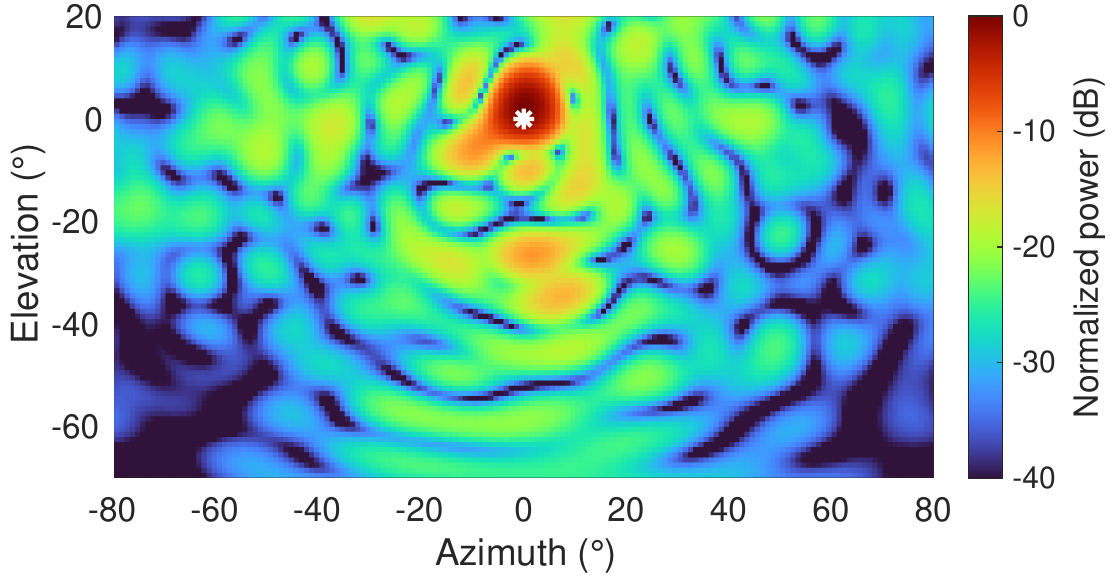}
{\footnotesize (a) Measured pattern: $\bar{\thetav} = [0^\circ,0^\circ]$}
\end{minipage}
\hfill
\begin{minipage}{0.32\linewidth}
\centering
\includegraphics[width=\linewidth]{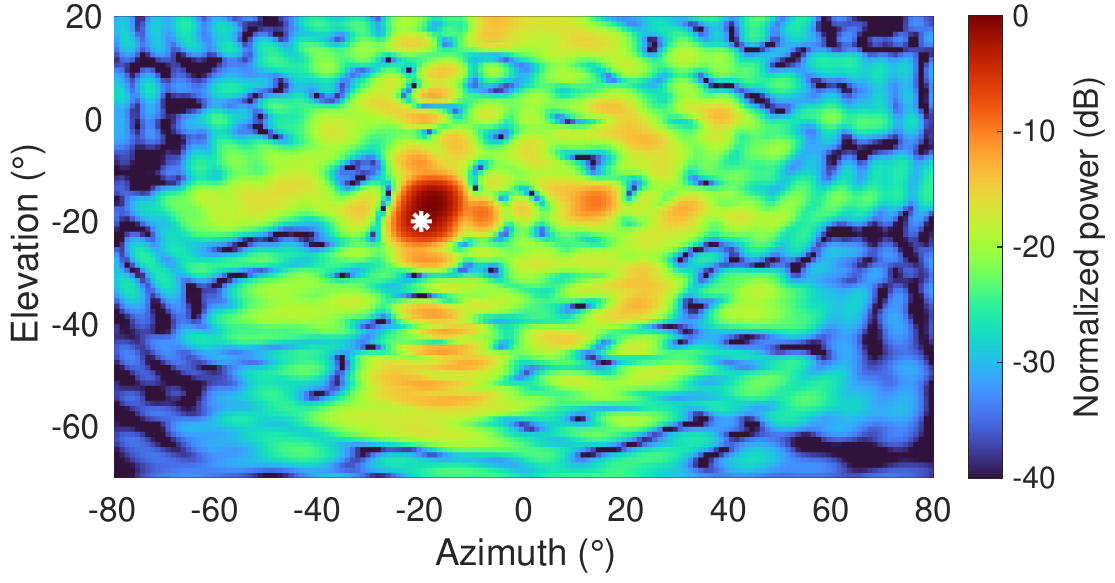}
{\footnotesize (b) Measured pattern: $\bar{\thetav} = [-20^\circ,-20^\circ]$}
\end{minipage}
\hfill
\begin{minipage}{0.32\linewidth}
\centering
\includegraphics[width=\linewidth]{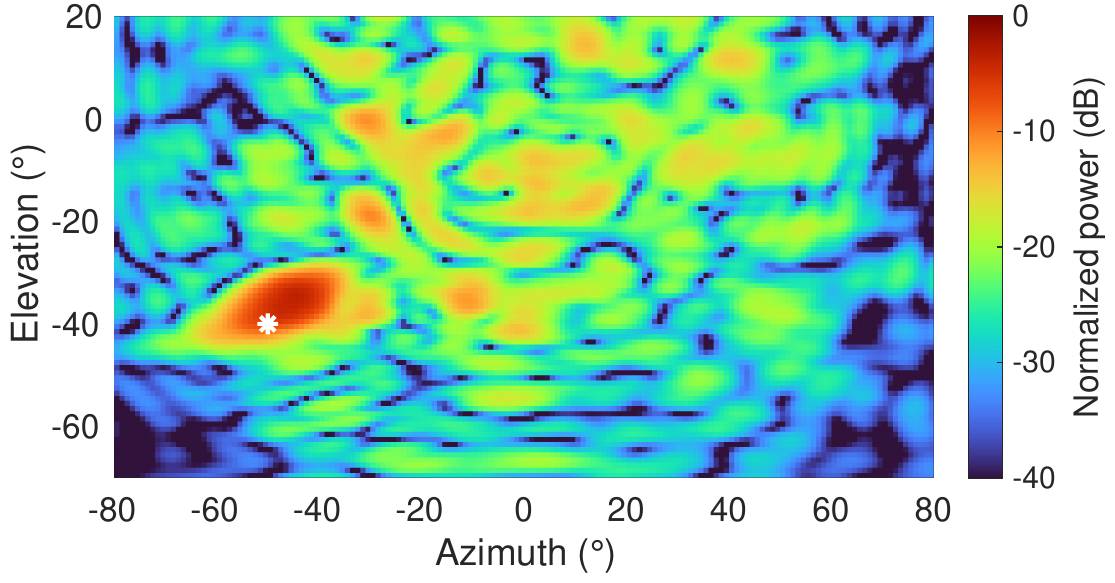}
{\footnotesize (c) Measured pattern: $\bar{\thetav} = [-50^\circ,-40^\circ]$}
\end{minipage}

\vspace{0.3cm}

\begin{minipage}{0.32\linewidth}
\centering
\includegraphics[width=\linewidth]{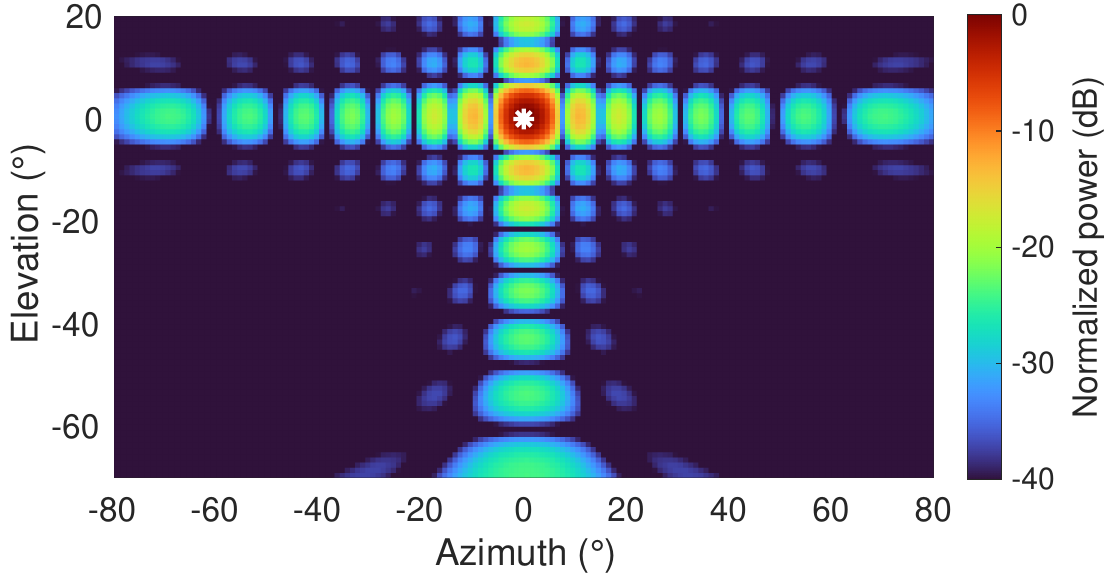}
{\footnotesize (d) Ideal pattern: $\bar{\thetav} = [0^\circ,0^\circ]$}
\end{minipage}
\hfill
\begin{minipage}{0.32\linewidth}
\centering
\includegraphics[width=\linewidth]{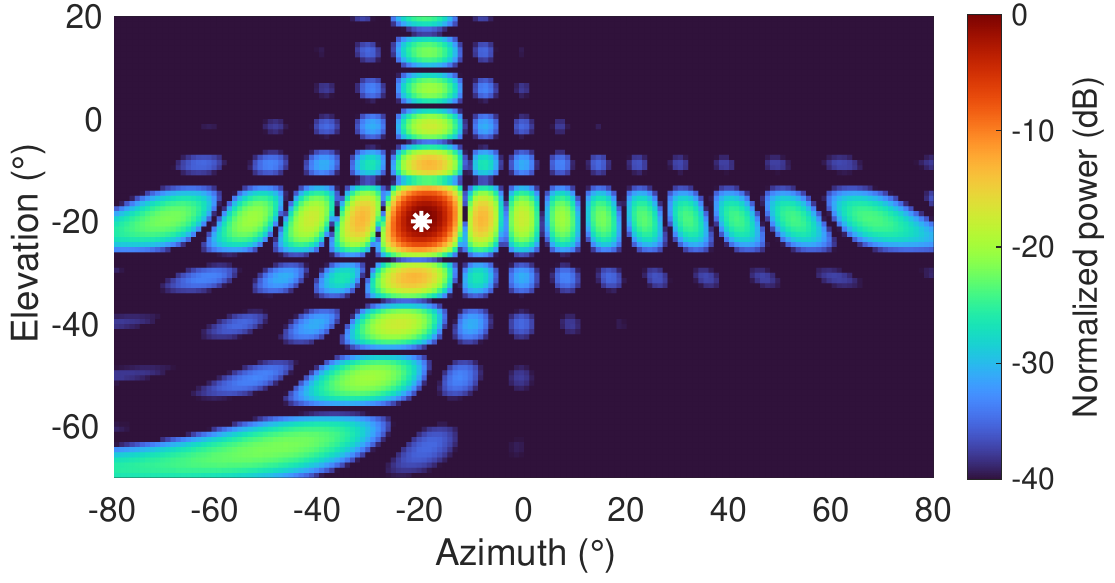}
{\footnotesize (e) Ideal pattern: $\bar{\thetav} = [-20^\circ,-20^\circ]$}
\end{minipage}
\hfill
\begin{minipage}{0.32\linewidth}
\centering
\includegraphics[width=\linewidth]{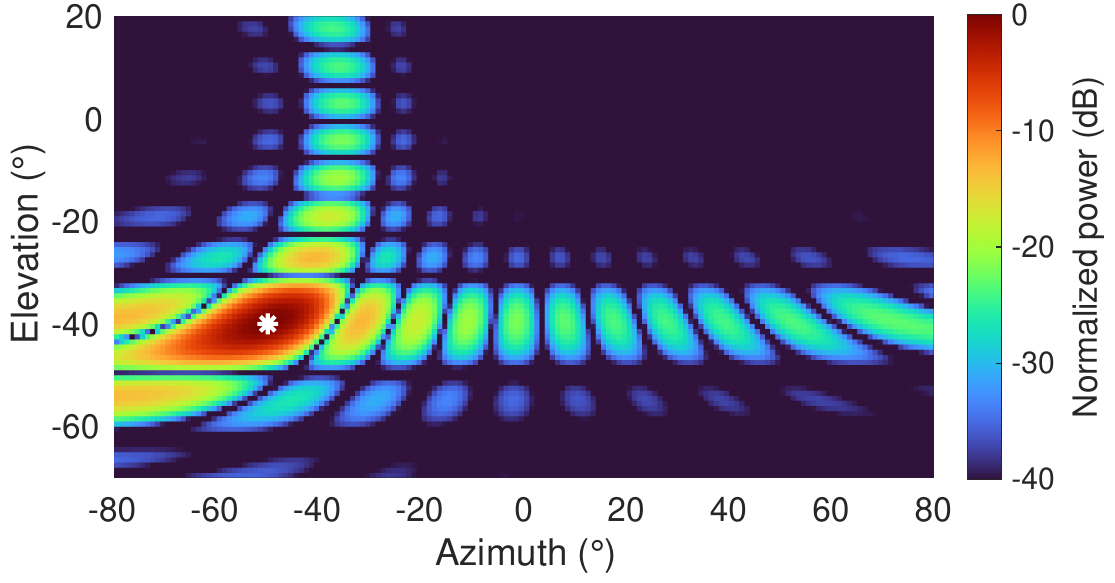}
{\footnotesize (f) Ideal pattern: $\bar{\thetav} = [-50^\circ,-40^\circ]$}
\end{minipage}

\caption{Visualization of the 3D RIS beam patterns with the measured patterns obtained at 5 GHz. The white asterisks shown in the figures indicate the preset scanning angles.}
\label{fig:beam_pattern_compare}

\end{figure*}




The measured RIS prototype is a low-cost single-layer reflective programmable metasurface, originally presented in~\cite{li2023design}. The RIS operates in the sub-6\,GHz band and employs 2-bit discrete phase tuning, providing four programmable phase states per element. The array comprises $16\times16$ elements, and the beamforming capability can be achieved through digital phase control.
The measurements were conducted in an anechoic chamber to characterize the intrinsic RIS beam response under different phase tuning configurations, free from environmental multipath effects, as shown in Fig.~\ref{fig:RISmea_setup}. 
Directional beamforming was considered in the measurements, where the reflected RIS beam was steered toward 66 predefined directions using corresponding phase tuning settings. These phase configurations constitute the codebook later employed in the RIS-aided positioning system. For each steering direction, the RIS beam pattern over the entire sphere was recorded with an angular resolution of $1^\circ$. The measurement settings are summarized in Table~\ref{tab:measurement_settings}.

Fig.~\ref{fig:beam_pattern_compare} compares the measured RIS beam patterns, i.e., the power of the ground-truth RIS beam response in the anechoic chamber, with the ideal beam response obtained using \eqref{eq:b_ideal} at different scanning angles $\bar\thetav = [\bar \theta_\text{azi}, \bar \theta_\text{ele}]$. As observed, the practical RIS beam responses obtained from measurements exhibit noticeable deviations from the ideal ones due to hardware imperfections and electromagnetic effects. First, the main beam direction and peak gain of the true RIS beam may shift from the preset steering angles, particularly at large scanning angles. Second, the sidelobe levels and their distributions differ significantly from those predicted by the ideal model, mainly due to mutual coupling among RIS elements and non-ideal element patterns. Third, the main lobe in true patterns tends to be less symmetric than in the ideal case, as imperfections caused by fabrication errors and non-ideal RF components may occur randomly across RIS elements. These distortions in true RIS beam patterns highlight the importance of RIS beam calibration for achieving high-accuracy positioning.

The proposed calibration framework assumes that the RIS beam pattern does not vary significantly within the bandwidth of interest, as discussed in Section~\ref{sec:signal_model}-A. 
\begin{figure}[t]
\centering
\includegraphics[width=0.7\linewidth]{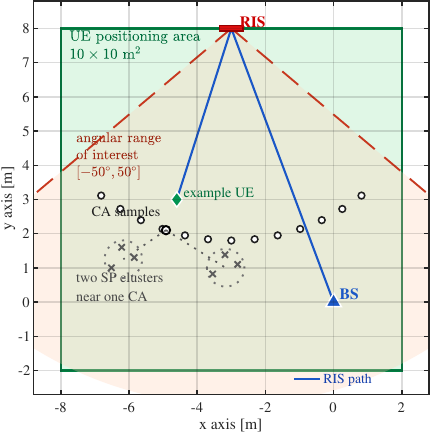}
\vspace{-0.15cm}
\caption{Top-view illustration of the numerical simulation scenario. The \ac{mpc} and \ac{los} paths are omitted for clarity.}
\label{fig:simulation_scenario}
\vspace{-0.3cm}
\end{figure}
\begin{figure}
\centering
\centerline{
\begin{tikzpicture}
\begin{axis}[
  width=\linewidth,
  height=0.58\linewidth,
  xmode=log,
  xmin=3.16227766017e-9,
  xmax=0.00001,
  ymin=0,
  ymax=1,
  xtick={1e-8,1e-7,0.000001,0.00001},
  ytick={0,0.2,0.4,0.6,0.8,1},
  grid=both,
  major grid style={draw=gray!35,line width=0.2pt},
  minor grid style={draw=gray!15,line width=0.1pt},
  tick label style={font=\footnotesize},
  label style={font=\footnotesize},
  xlabel={$\epsilon$ },
  ylabel={$\Pr(e_{g,s}^{\mathrm{R}}<\epsilon)$},
  axis line style={black,line width=0.45pt},
  tick style={black,line width=0.45pt},
  legend cell align={left},
  ylabel style={yshift=-1mm},
  legend columns=2,
  legend style={at={(0.5,-0.24)},anchor=north,font=\footnotesize,draw=none,fill=none,row sep=0.3pt,/tikz/every even column/.append style={column sep=0.8em}}
]
\addplot+[color={rgb,1:red,0;green,0;blue,0}, dashed, line width=0.9pt, mark=none] table[x=x,y=y] {figures/tikz_data/CDF_mpc_bw_01.dat};
\addlegendentry{No MPCs, 100MHz}
\addplot+[color={rgb,1:red,0.3;green,0.3;blue,0.3}, solid, line width=0.9pt, mark=square, mark size=1.7pt, mark repeat=33, mark phase=1, mark options={solid,fill=white,line width=0.55pt}] table[x=x,y=y] {figures/tikz_data/CDF_mpc_bw_02.dat};
\addlegendentry{12 MPCs, $\sigma=0.5$, 100MHz}
\addplot+[color={rgb,1:red,0;green,0.447;blue,0.741}, solid, line width=0.9pt, mark=diamond, mark size=1.7pt, mark repeat=33, mark phase=1, mark options={solid,fill=white,line width=0.55pt}] table[x=x,y=y] {figures/tikz_data/CDF_mpc_bw_03.dat};
\addlegendentry{16 MPCs, $\sigma=0.5$, 100MHz}
\addplot+[color={rgb,1:red,0.85;green,0.325;blue,0.098}, solid, line width=0.9pt, mark=triangle, mark size=1.7pt, mark repeat=33, mark phase=1, mark options={solid,fill=white,line width=0.55pt}] table[x=x,y=y] {figures/tikz_data/CDF_mpc_bw_04.dat};
\addlegendentry{16 MPCs, $\sigma=2$, 100MHz}
\addplot+[color={rgb,1:red,0.466;green,0.674;blue,0.188}, solid, line width=0.9pt, mark=triangle, mark size=1.7pt, mark repeat=33, mark phase=1, mark options={solid,fill=white,line width=0.55pt,rotate=180}] table[x=x,y=y] {figures/tikz_data/CDF_mpc_bw_05.dat};
\addlegendentry{12 MPCs, $\sigma=0.5$, 60MHz}
\addplot+[color={rgb,1:red,0.494;green,0.184;blue,0.556}, solid, line width=0.9pt, mark=asterisk, mark size=1.7pt, mark repeat=33, mark phase=1, mark options={solid,fill=white,line width=0.55pt}] table[x=x,y=y] {figures/tikz_data/CDF_mpc_bw_06.dat};
\addlegendentry{12 MPCs, $\sigma=0.5$, 140MHz}
\end{axis}
\end{tikzpicture}}
\vspace{-0.1cm}
\caption{CDF of the RIS channel estimation error for different multipath environments and frequency bandwidths.}
\label{fig:CDF_mpc_bw}
\vspace{-0.2cm}
\end{figure}
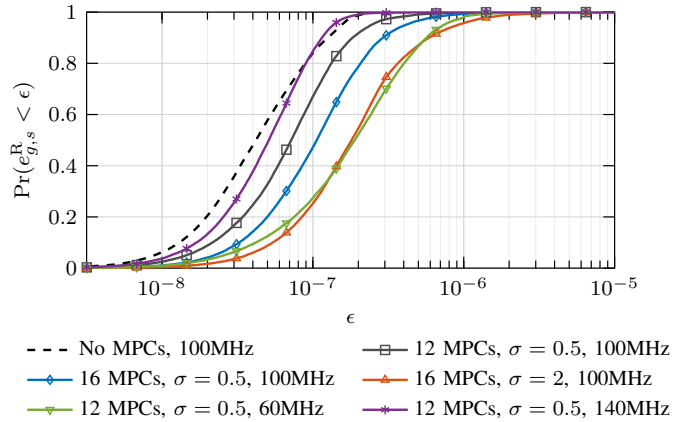
\begin{figure*}[t]
    \centering

    \begin{minipage}[t]{0.32\textwidth}
        \centering
        \includegraphics[width=\linewidth]{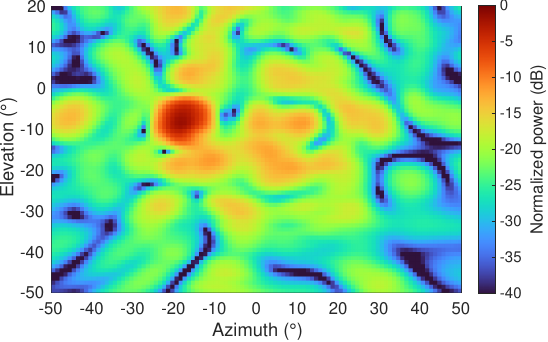}
        \vspace{1mm}
        \parbox{\linewidth}{\centering\scriptsize (a) Calibrated RIS beam, $1^\circ$ sampling}
    \end{minipage}
    \hfill
    \begin{minipage}[t]{0.32\textwidth}
        \centering
        \includegraphics[width=\linewidth]{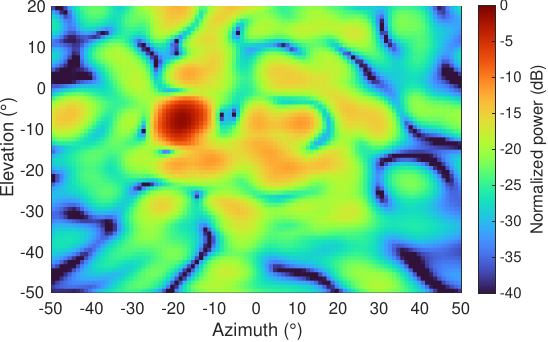}
        \vspace{1mm}
        \parbox{\linewidth}{\centering\scriptsize (b) Calibrated RIS beam, $2^\circ$ sampling}
    \end{minipage}
    \hfill
    \begin{minipage}[t]{0.32\textwidth}
        \centering
        \includegraphics[width=\linewidth]{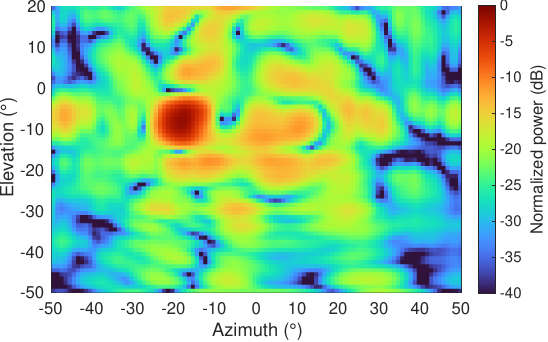}
        \vspace{1mm}
        \parbox{\linewidth}{\centering\scriptsize (c) Calibrated RIS beam, $4^\circ$ sampling}
    \end{minipage}

    \vspace{0.18cm}

    \makebox[\textwidth][c]{%
        \begin{minipage}[t]{0.32\textwidth}
            \centering
            \includegraphics[width=\linewidth]{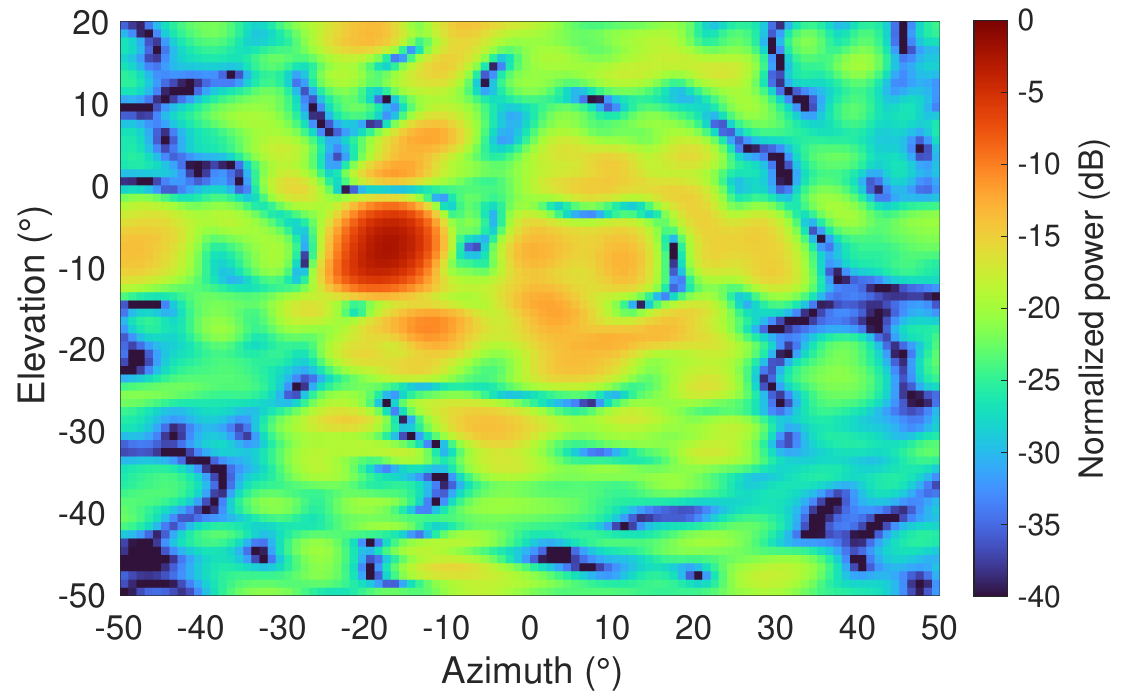}
            \vspace{1mm}
            \parbox{\linewidth}{\centering\scriptsize (d) Ground truth}
        \end{minipage}
        \hspace{0.06\textwidth}
        \begin{minipage}[t]{0.32\textwidth}
            \centering
            \includegraphics[width=\linewidth]{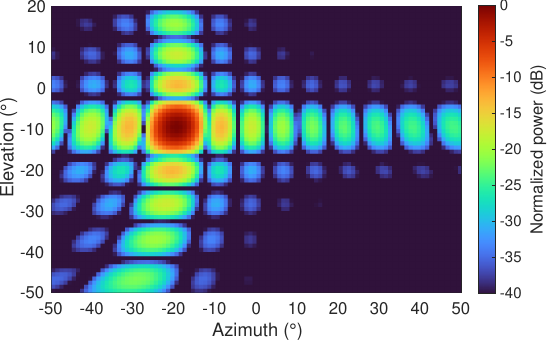}
            \vspace{1mm}
            \parbox{\linewidth}{\centering\scriptsize (e) Ideal RIS beam}
        \end{minipage}%
    }
    \caption{Visualization of RIS beam response power. Panels (a)--(c) show the reconstructed calibrated beams with angular sampling steps of $1^\circ$, $2^\circ$, and $4^\circ$, respectively, whose average BRS values over 66 codewords are $88.5\%$, $87.3\%$, and $85.2\%$. Panels (d) and (e) show the ground truth beam and the ideal RIS beam, respectively; the ideal beam response model achieves an average BRS of only $43.7\%$.}

    \label{fig:BRS_compare}
    \vspace{-0.2cm}
\end{figure*}
\subsection{Integration into Numerical Simulations}
The measured complex 3D beam patterns were incorporated into the numerical simulation framework to generate the received signals at the CA/UE side by replacing the ideal beam model with the measured beam responses. Specifically, for each predefined phase configuration of RIS, the measured beam response was interpolated onto the angular grid used in the simulations.
This measurement-driven modeling enables realistic evaluation of:
\begin{itemize}
    \item The effectiveness of the proposed on-site beam calibration algorithm in terms of RIS channel estimation, and the calibrated RIS beam response.
    \item The impact of RIS beam calibration on positioning accuracy using ALB.
\end{itemize}
We consider a 3D RIS-aided positioning system with one UE and one \ac{ris}. The simulation parameters are summarized in Table~\ref{tab:settings}. The CA is deployed to cover the RIS beam response within the angular ranges $[-50^\circ, 50^\circ]$ and $[-50^\circ, 10^\circ]$ in the azimuth and elevation domains, respectively. Fig.~\ref{fig:simulation_scenario} provides a top-view illustration of the numerical simulation geometry.

As shown in Section~\ref{sec:algorithm}-A, the first step of the calibration algorithm is to extract the channel responses corresponding only to the RIS-reflected path. To evaluate the impact of multipath propagation, we consider two clusters of \acp{sp}, both located near the CA. The distances between the centers of the clusters and the CA are set to 1~m and 2~m, respectively. In the default setting, each cluster contains three \acp{sp} uniformly distributed within a disk of radius 1~m on the $xoy$ plane, resulting in six single-bounce \acp{mpc} and six RIS-associated \acp{mpc} in total. All \acp{mpc} are assigned identical \ac{rcs} coefficients. 
The channel estimation error of the RIS path is evaluated using \(e_{g,s}^{\mathrm{R}}\) in~\eqref{eq:channel_squared_error}, which measures the squared error between the estimated and true entries of the RIS beam response matrix.
The empirical CDF is then computed from all \(GS\) error samples \(\{e_{g,s}^{\mathrm{R}}\}_{g=1,s=1}^{G,S}\). 

Fig.~\ref{fig:CDF_mpc_bw} presents the resulting CDFs under different multipath conditions and signal bandwidths. In addition, we consider a denser multipath scenario in which each cluster contains four \acp{sp}, resulting in 16 \acp{mpc} in total.
Compared with the case without \acp{mpc}, the estimation error increases when \acp{mpc} are introduced, as multipath components with delays close to that of the RIS-reflected path cannot be well separated under limited bandwidth. Increasing the number of \acp{mpc} from 12 to 16 further degrades the estimation accuracy. Moreover, a larger \ac{rcs}, e.g., $\sigma_{\mathrm{RCS}} = 2\,\mathrm{m}^2$, produces stronger multipath interference and consequently larger estimation errors. For the case with 12~\acp{mpc} and $\sigma_{\mathrm{RCS}} = 0.5\,\mathrm{m}^2$, the results obtained with bandwidths of 60, 100, and 140~MHz show that increasing the bandwidth improves the estimation accuracy, owing to the enhanced delay resolution and better separation of propagation paths. Overall, these results demonstrate that the proposed calibration method remains robust under moderate multipath conditions, while larger bandwidths help improve estimation accuracy.

In the second step of the calibration algorithm, the parameters of the RIS beam response model in \eqref{eq:b_proposed2} are estimated. After calibration, the RIS beam response can be reconstructed from the calibrated model. Fig.~\ref{fig:BRS_compare} compares the ground-truth RIS beam response with the reconstructed responses after calibration and the ideal beam response for the scanning angle $\bar{\thetav} = [-20^\circ, -10^\circ]$. For the calibrated beams obtained with reduced samples, i.e., with an angular sampling step larger than $1^\circ$, interpolation is applied only for visualization and to facilitate comparison. It can be observed that the main beam and high-power sidelobes are accurately reconstructed after calibration. In contrast, some discrepancies appear in the low-power sidelobes, particularly at larger angular offsets. The average BRS values\footnote{The BRS values are evaluated on a dense angular grid with \(1^\circ\) resolution over \([-50^\circ,50^\circ]\) in azimuth and \([-50^\circ,20^\circ]\) in elevation, rather than only on the calibration sample points.} over 66 codewords for the calibrated beam responses with angular sampling steps of \(1^\circ\), \(2^\circ\), and \(4^\circ\) are 88.5\%, 87.3\%, and 85.2\%, respectively, whereas the average BRS for the ideal beam response model is only 43.7\%. Denser sampling leads to a slight improvement in the reconstructed beam response. Nevertheless, the calibrated RIS beam responses obtained with different sampling steps all achieve high average BRS values, demonstrating the effectiveness of the proposed calibration method.


\begin{table}[t]
\centering
\caption{Default Simulation Settings for Calibration and Positioning}
\footnotesize
\setlength{\tabcolsep}{3pt}
\renewcommand{\arraystretch}{1.25}
\begin{tabular}{|p{0.34\linewidth}|p{0.56\linewidth}|}
\hline
\multicolumn{2}{|c|}{\textbf{Signal \& Frequency Parameters}} \\ \hline
Center frequency, $f_c$ & $5~\mathrm{GHz}$ \\ \hline
Bandwidth, BW & $100~\mathrm{MHz}$ \\ \hline
Number of subcarriers & $K=128$ \\ \hline
Transmit power & $\text{30}~\mathrm{dBm}$ \\ \hline
Noise figure & $\mathrm{NF}=\text{10}~\mathrm{dB}$ \\ \hline
\multicolumn{2}{|c|}{\textbf{RIS, Calibration, and Positioning Geometry}} \\ \hline
BS position & $\pv_B=[0,0,0]^\top~\mathrm{m}$ \\ \hline
RIS position & $\pv_R=[-3,8,0]^\top~\mathrm{m}$ \\ \hline
RIS orientation & $\ov_R=[\alpha_R,\beta_R,\gamma_R]^\top=[\text{0},\text{90},\text{0}]^\top$ \\ \hline
RIS dimensions & $N=N_r\times N_c=16\times16$ \\ \hline
RIS codebook & $66$ preset steering directions \\ \hline
CA angular coverage & Az. $[-50^\circ,50^\circ]$; el. $[-50^\circ,10^\circ]$ \\ \hline
CA angular sampling step & $1^\circ$ \\ \hline
UE positioning area & $x_u \in [-8,2]$, $y_u \in [-2,8]$, $z_u=0$ (m) \\ \hline
\multicolumn{2}{|c|}{\textbf{Multipath Parameters}} \\ \hline
Number of clusters & 2 near the CA \\ \hline
\acp{sp} per cluster & 3 \\ \hline
\ac{sp} distribution & Uniform within a disk of radius $1~\mathrm{m}$ on the $xoy$ plane \\ \hline
Number of \acp{mpc} & $M+Q=6+6=12$ \\ \hline
\ac{rcs} coefficient, $\sigma_{\mathrm{RCS}}$ & $0.5~\mathrm{m}^{2}$ \\ \hline
\end{tabular}
\label{tab:settings}
\end{table}


\begin{figure}[t]
\centering

\begin{minipage}{0.9\linewidth}
\centering
\includegraphics[width=0.95\linewidth]{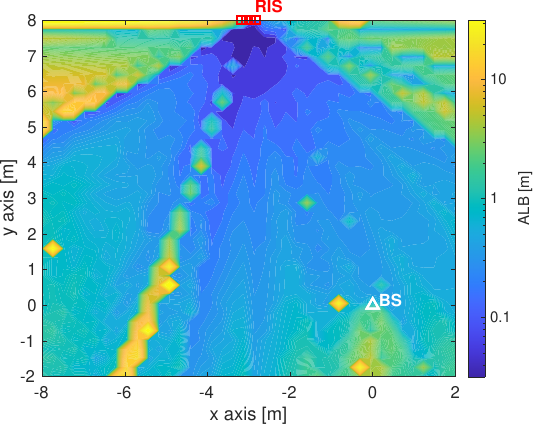}
\centerline{\small (a) Without calibration}
\end{minipage}

\vspace{0.2cm}

\begin{minipage}{0.9\linewidth}
\centering
\includegraphics[width=0.95\linewidth]{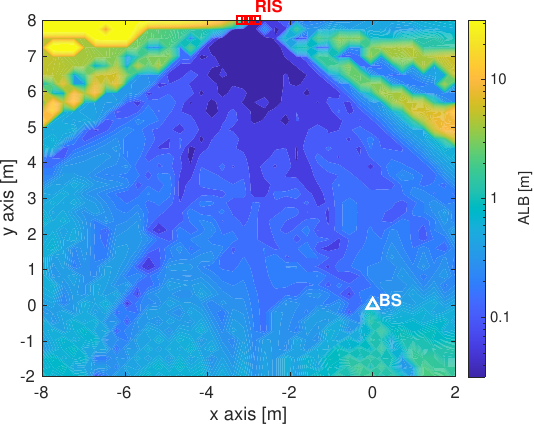}
\centerline{\small (b) With calibration}
\end{minipage}

\caption{Visualization of ALB when the UE moves within a $10\times10$\,m$^2$ area while the BS and RIS remain fixed.}
\label{fig:ALB_compare}

\end{figure}

\begin{figure}
\centering
\centerline{
\begin{tikzpicture}
\begin{axis}[
  width=\linewidth,
  height=0.58\linewidth,
  xmode=log,
  xmin=0.01,
  xmax=15.8489319246,
  ymin=0,
  ymax=1,
  xtick={0.01,0.1,1,10},
  ytick={0,0.2,0.4,0.6,0.8,1},
  grid=both,
  major grid style={draw=gray!35,line width=0.2pt},
  minor grid style={draw=gray!15,line width=0.1pt},
  tick label style={font=\footnotesize},
  label style={font=\footnotesize},
  xlabel={$\epsilon$ [m]},
  ylabel={$\Pr(\mathrm{ALB}<\epsilon)$},
  axis line style={black,line width=0.45pt},
  tick style={black,line width=0.45pt},
  legend cell align={left},
  legend columns=2,
  legend style={at={(0.03,0.97)},anchor=north west,font=\footnotesize,draw=black,fill=white,fill opacity=0.9,text opacity=1,row sep=0.5pt,/tikz/every even column/.append style={column sep=0.8em}}
]
\addplot+[color={rgb,1:red,0;green,0;blue,0}, solid, line width=0.95pt, mark=o, mark size=2.1pt, mark repeat=67, mark phase=1, mark options={solid,fill=white,line width=0.55pt}] table[x=x,y=y] {figures/tikz_data/ALB_CDF_01.dat};
\addlegendentry{Ideal}
\addplot+[color={rgb,1:red,0;green,0.447;blue,0.698}, dashed, line width=0.95pt, mark=square, mark size=2.1pt, mark repeat=67, mark phase=18, mark options={solid,fill=white,line width=0.55pt}] table[x=x,y=y] {figures/tikz_data/ALB_CDF_02.dat};
\addlegendentry{Cal., $1^\circ$}
\addplot+[color={rgb,1:red,0.835;green,0.369;blue,0}, dash dot, line width=0.95pt, mark=triangle, mark size=2.1pt, mark repeat=67, mark phase=35, mark options={solid,fill=white,line width=0.55pt}] table[x=x,y=y] {figures/tikz_data/ALB_CDF_03.dat};
\addlegendentry{Cal., $2^\circ$}
\addplot+[color={rgb,1:red,0;green,0.62;blue,0.451}, dotted, line width=0.95pt, mark=diamond, mark size=2.1pt, mark repeat=67, mark phase=52, mark options={solid,fill=white,line width=0.55pt}] table[x=x,y=y] {figures/tikz_data/ALB_CDF_04.dat};
\addlegendentry{Cal., $4^\circ$}
\end{axis}
\end{tikzpicture}}
\vspace{-0.1cm}
\caption{CDF of the ALB when the UE moves within a $10\times10$\,m$^2$ area while the BS and RIS remain fixed.}
\label{fig:ALB_CDF}
\vspace{-0.2cm}
\end{figure}
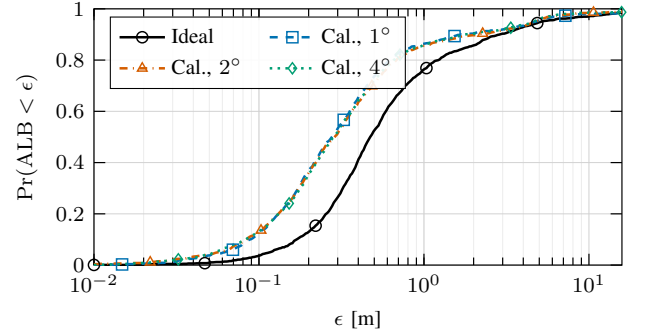
To directly evaluate the impact of calibration on RIS-aided positioning, we use the ALB metric defined in Section~\ref{sec:metrics}-B. The ALB is computed for fixed \ac{bs} and \ac{ris} locations, while the \ac{ue} position is swept over a $10 \times 10$~m$^2$ area. Fig.~\ref{fig:ALB_compare} compares the ALB with and without calibration. Without calibration, the ideal beam model in \eqref{eq:b_ideal} is used for positioning, whereas with calibration, the calibrated beam model in \eqref{eq:b_proposed2} is used with an angular sampling step of $1^\circ$. Large ALB values are observed outside the calibrated beam-scanning range in both cases, since these locations are outside the angular range of interest and receive low RIS-reflected signal power. Within the calibrated angular range, the ALB mainly varies with the \ac{aod} from RIS to UE. For the case with calibration, slightly higher ALB values are observed near gaps between adjacent scanning directions. Without calibration, localized high-ALB regions appear where the ideal beam model mismatch is more pronounced and the received power is low. Calibration significantly reduces these high-ALB regions, demonstrating its effectiveness in improving positioning accuracy. 

The corresponding \acp{cdf}, including the results obtained with reduced calibration samples, are shown in Fig.~\ref{fig:ALB_CDF}. Calibration consistently shifts the ALB distribution toward smaller values. In particular, the probability that the ALB is below $0.5~\mathrm{m}$ increases from 0.52 without calibration to 0.74 with an angular sampling step of $1^\circ$, and remains around 0.72 for sampling steps of $2^\circ$ and $4^\circ$. The small differences among the calibrated curves indicate that reduced angular sampling still preserves most of the improvement in positioning accuracy. These results confirm that the proposed calibration method effectively mitigates RIS beam model mismatch and improves the positioning accuracy of the RIS-aided localization system.

\section{Conclusion}
\label{sec:conclusion}
An on-site RIS beam calibration framework is proposed in this work to improve the positioning performance of RIS-aided localization systems. The presented RIS beam model captures several key non-idealities and is validated using measured beam patterns from a practical RIS prototype. The proposed calibration algorithm mitigates multipath effects by exploiting a delay-domain sparse recovery approach and estimates the required beam model parameters using a gradient-descent-based method. Notably, the calibration process does not require adjusting the RIS phase tuning network and can be performed under the preset/default RIS beamforming codebook. Furthermore, the proposed method is scalable to large-scale RIS deployments and can directly utilize standard 5G OFDM signals, making it a promising solution for on-site RIS beam calibration in future radio systems. Nevertheless, this work focuses on the far-field scenario. Extending the framework to near-field calibration of antenna arrays/RIS and exploring machine-learning-based calibration algorithms to reduce computational complexity constitute important directions for future research.  

\bibliographystyle{IEEEtran}
\bibliography{main}

\end{document}